\documentclass[reprint,twocolumn,superscriptaddress,showkeys,
nofootinbib,notitlepage,amsmath,amssymb,floatfix]{revtex4-2}

\usepackage{latexsym,amsmath,amssymb,lmodern,float,url}
\usepackage{natbib}
\usepackage{color}
\usepackage{multirow}
\usepackage{bm}
\usepackage{array}
\usepackage[normalem]{ulem}
\usepackage{cleveref}
\usepackage{xcolor}
\usepackage{tabularx}

\usepackage{mathtools}

\def\de{\delta^{\vphantom{1}}}
\def\bde{{\bar\delta}}

\def\h3{{\displaystyle{\frac 3 2}}}

\newcommand{\bbar}{\overline}

\def\schro{Schr\"odinger~}

\begin{document}
\title{Spin Dependence of Meson Thresholds in the Diabatic Dynamical Diquark Model}
\author{Richard F. Lebed}
\email{Richard.Lebed@asu.edu}
\affiliation{Department of Physics, Arizona State University, Tempe,
AZ 85287, USA}
\date{July, 2025}

\begin{abstract}
The diabatic dynamical diquark model is designed to unify diquark and molecular approaches for exotic hadrons, by including the effects of di-meson thresholds on fundamental diquark-antidiquark states in order to form physical tetraquarks.  We generalize this model by implementing the consequences of heavy-quark spin symmetry for each di-meson pair.  We include the specific spin factors and mass differences associated with mesons related by this symmetry (such as $D, D^*$), both features being necessary for the systematic incorporation of effects beyond the static, heavy-source limit inherent to the (adiabatic) Born-Oppenheimer treatment.  We obtain explicit interaction potentials for all $J = 0, 1, 2$ states, along the way deriving unusual variants of Fierz reordering identities, and discuss the application of the potentials to specific physical states such as $\chi_{c1}(3872)$ and $T_{c\bar c 1}(3900)$.
\end{abstract}

\maketitle

\section{Introduction}\label{sec:Intro}

The pace of discovery in the field of heavy-quark exotic hadrons [{\it i.e.}, any hadron containing a significant component that is neither a valence quark-antiquark pair $q\bar q$ nor a quark (or antiquark) trio $qqq$] over the past two decades has been nothing less than breathtaking.    As late as the early 2000s, one could still argue that some unknown dynamical mechanism of QCD restricts all clearly observable physical hadronic states to appear solely in conventional $q\bar q$ and $qqq$ configurations.  Just over two decades later, at least 70 heavy-quark ($c,b$) states that are not easily explained as conventional hadrons have been observed~\cite{ParticleDataGroup:2024cfk}, spread over several flavor sectors: hidden-charm (both nonstrange and strange, both tetraquarks and pentaquarks), hidden-bottom, open-charm/strange, and double open-charm.

A particularly interesting property of the exotic candidates is that many of them lie close, sometimes extraordinarily so, to di-hadron thresholds, suggesting a substantial role for the di-hadron state as a constituent in the structure of the full exotic hadron.  The most spectacular example of this phenomenon is perhaps the first heavy-quark exotic candidate discovered, $\chi_{c1}(3872)$.  Using the latest mass measurements~\cite{ParticleDataGroup:2024cfk},
\begin{equation} \label{eq:Xbind}
m_{\chi_{c1}(3872)} - m_{D^0} - m_{D^{*0}} = -0.05 \pm 0.09 \ {\rm MeV} \, ,
\end{equation}
implying that the di-meson state $D^0 \bar D^{*0}$ (understood to include its charge conjugate) is undoubtedly a prominent component of $\chi_{c1}(3872)$.  And yet, multiple heavy-quark exotic candidates lie many MeV from plausible constituent di-hadron thresholds, such as the charmoniumlike $J^{PC} \! = \! 1^{--}$ states in excess of conventional $c\bar c$ states $\psi$ [{\it e.g.}, $\psi(4230)$, previously $Y(4230)$], and the $\pi^+ \! \psi(2S)$ resonance $T_{c\bar c 1}(4430)^+$ [previously $Z_c(4430)$].  In such cases, a diquark-antidiquark ($\de\bde$) structure can provide an alternative paradigm for explaining the existence of a spectrum of isolated tetraquark states.

Typical diquark models suffer a significant drawback by not explaining why a $\de\bde$ structure should be favored over the (presumably more stable) bound di-hadron configuration.  However, di-meson thresholds occupy only a fraction of the space of total energy values available to the 4-quark states, thus allowing substantial regions of energy in which $\de\bde$ structures are preferred.  Moreover, even for energy values not far from such thresholds, the 4-quark state may form in a configuration in which the two quarks $qq$ or the two antiquarks $\bar q \bar q$ lie closer to each other than either of the $q$ lies to either of the $\bar q$; and since the short-distance attraction of a $qq$ pair in its combined color-$\bar{\bf 3}$ configuration is fully half as large as the corresponding $q\bar q$ color-$\bf 1$ attraction, the $qq$ and $\bar q\bar q$ attractions can produce a $\delta_{\bf \bar 3}\bde^{\vphantom{+}}_{\bf 3}$ configuration that persists---under the assumption that these attractions support the continued presence of $\de\bde$ state content---even in energy regions not far from di-meson thresholds.

These observations form the essence of the {\it dynamical diquark picture}~\cite{Brodsky:2014xia}.  It can also be extended to pentaquarks~\cite{Lebed:2015tna} by noting that the aggregation of quarks through successive combination of color-{\bf 3} quarks allows one to model the pentaquark as a $(qq)_{\bar{\bf 3}} {[\bar q_{\bar{\bf 3}} (q q)_{\bar{\bf 3}}]^{\vphantom{}}_{\bf 3}}$ (a diquark-{\it triquark}) compound.  This picture is elevated to the predictive {\it dynamical diquark model}~\cite{Lebed:2017min} in cases in which each diquark (or triquark) quasiparticle contains a heavy quark, hence allowing for the configuration to develop into that of two heavy, slow sources connected by a color flux tube, ideally suited for use in the Born-Oppenheimer (BO) approximation.  The interaction provided by the flux tube is the $\bf 3$-$\bar{\bf 3}$ static potential calculated in multiple lattice-QCD simulations ({\it e.g.}, Refs.~\cite{Morningstar:2019,Capitani:2018rox}), typically to study the spectra of conventional and hybrid heavy-quark mesons.

The dynamical diquark model has been thoroughly explored in a series of papers that present the results of numerical analysis in multiple flavor sectors and state configurations~\cite{Giron:2019bcs,Giron:2019cfc,Giron:2020fvd,Giron:2020qpb,Giron:2020wpx,Giron:2021sla,Giron:2021fnl,Jafarzade:2025qvx}, but relying (in these works) exclusively upon the {\it adiabatic\/} heavy-source limit inherent to the BO approximation.  However, the composition of the physical mass eigenstate necessarily changes rapidly in the energy region near di-hadron thresholds; for example, the $\de\bde$ system with a mass near the $D^0 \bar D^{*0}$ threshold may form the kernel of the state that ultimately becomes $\chi_{c1}(3872)$, but the $D^0 \bar D^{*0}$ state dominates the composition of the physical $\chi_{c1}(3872)$ state.  This {\it diabatic\/} extension of the dynamical diquark model~\cite{Lebed:2022vks,Lebed:2023kbm,Lebed:2024rsi,Lebed:2025xbz}, which descibes the mechanism of this configuration mixing, is the formalism in which the calculations of this paper are performed; it is outlined in the next section.

Detailed calculations performed thus far in the diabat\-ic dynamical diquark model assume a universal interac\-tion between the fundamental $\de\bde$ state and threshold di-meson pair.  Of course, such interactions should also depend upon the spin and flavor quantum numbers of both the fundamental and the hadronic state.  In the case of hadrons containing a heavy quark $Q$, couplings for states that differ only through the orientation of the spin of $Q$ (such as $D$ and $D^*$) are related by heavy-quark spin symmetry (HQSS)~\cite{Manohar_Wise_2023}.  The diabatic mixing of conventional heavy quarkonium ($Q\bar Q$) states with (universal) di-meson thresholds was first described in Ref.~\cite{Bruschini:2020voj}, and the effects obtained from imposing HQSS were derived in Ref.~\cite{Bruschini:2023zkb}.  Our purpose is to extend this HQSS formalism to the mixing of di-meson thresholds with $\de\bde$ states.  This derivation requires innovations such as the incorporation of states with larger spin eigenvalues and the generalization of the best-known Fierz reordering theorem for 4-fermion operators.  As an illustration of the results, we present explicit expressions for diabatic couplings of the lowest  well-known hidden-charm exotics: $\chi_{c1}(3872)$, $T_{c\bar c 1}(3900)$ [formerly $Z_c(3900)$], and $T_{c\bar c 1}(4020)$ [formerly $Z_c(4020)$].  Note that the latter two states have $I \! = \! 1$ and thus cannot mix with conventional $c\bar c$ states.

This paper is structured as follows.  In Sec.~\ref{sec:Model} we review the dynamical diquark model briefly in its original (adiabatic) form, and more thoroughly in its diabatic extension.  Section~\ref{sec:Thresh} reviews the key points in the constraints of HQSS upon heavy-meson thresholds and applies them to the $\de\bde$ system, which is where the generalized Fierz reorderings become crucial.  Applications to specific low-lying hidden-charm states appear in Sec.~\ref{sec:Examples}.   We then summarize our findings and describe future prospects of this model in Sec.~\ref{sec:Concl}.

\section{The Diabatic Dynamical Diquark Model} \label{sec:Model}

\subsection{Original Adiabatic Model}

The original dynamical diquark model may be described as {\it adiabatic\/} because it employs the Born-Oppen\-heimer (BO) approximation between two heavy (and hence slowly moving) diquark sources, while the lighter degrees of freedom (d.o.f.) comprising the gluon flux tube {({\it i.e.}, gluons and sea-quark pairs, but not the light valence quarks)} connecting them adapts quickly (adiabatically) to their motion.   Intrinsic to this interpretation is that each diquark must contain a heavy quark $Q$ ($c$ or $b$) or antiquark, although the marginal case with $s$ quarks has also been explored~\cite{Jafarzade:2025qvx}.  We define as usual the diquark $\de \equiv ( Q q )$ and antidiquark $\bde \equiv ( \bar Q \bar q )$ quasiparticles, where $q$ is a light quark ($u,d$) whose flavor can differ between $\de$ and $\bde$ in order to produce both $I \! = \! 0$ and $I \! = \! 1$ states.  $\de$ and $\bde$ are assumed to transform as color-$\bar{\bf 3}$ and -${\bf 3}$ combinations, respectively, which unlike the alternate color-${\bf 6}$ and -$\bar{\bf 6}$ possibilities, are attractive at small internal $Qq$ and $\bar Q \bar q$ separations.

The static ${\bf 3}$-$\bar{\bf 3}$ BO potential $V(r)$ connecting two such heavy sources (and hence describing the state of the light d.o.f.) is the same one as appearing in heavy quarkonium or its hybrid states, and has been calculated in lattice QCD simulations for decades; for recent determinations, see Refs.~\cite{Morningstar:2019,Capitani:2018rox}.  Since all of these systems exhibit a cylindrical ($D_{\infty h}$) rather than a spherical [SO(3)] symmetry, the appropriate quantum numbers for describing the configuration of the BO potential are ones that have been used for many decades in the context of diatomic molecules ({\it e.g.}, Ref.~\cite{Landau:1977}).  Given the orbital angular momentum ${\bf J}_{\rm light}$ of the light d.o.f., one begins with the unit vector $\hat{\bm{r}}$ connecting $\bde$ to $\de$ (separation $r$) and chooses any plane ${\cal P}$ containing $\hat{\bm{r}}$.  The configuration symbol is then $\Lambda^\epsilon_\eta$, where $\Lambda \equiv | {\bf J}_{\rm light} \cdot \hat{\bm{r}}|$ and is denoted as $\Sigma$, $\Pi$, $\Delta$, $\ldots$ for $\Lambda = 0, 1, 2, \ldots$; $\epsilon$ is the eigenvalue of the light d.o.f.\ wave function upon reflection through ${\cal P}$; and $\eta$ (denoted by $g,u$ for $+1$, $-1$, respectively) is the light-d.o.f.\ $CP$ eigenvalue, where $P$ indicates spatial inversion through the $\bde$-$\de$ midpoint.  In this nomenclature, the lattice-QCD simulations described above have numerically determined that the potentials $V_{\Lambda^\epsilon_\eta} (r)$ in order of increasing energy are $\Sigma^+_g$, $\Pi_u^+$, and $\Sigma^-_u$.  Within these BO potentials, the states occupy orbitals labeled by principal $n$ and orbital $L$ quantum numbers; combining them with the quantum numbers of the heavy sources $\de,\bde$ gives the complete spectrum of tetraquark states (or with a heavy triquark source $\bar \theta$, gives the complete spectrum of pentaquark states) in the dynamical diquark model~\cite{Lebed:2017min}.

In fact, the known exotic states in this model appear to occupy only the ground-state BO potential $\Sigma^+_g (r)$, whose lattice determination, the same as for $Q\bar Q$, is fit well by a Cornell-like form:
\begin{equation}
\label{eq:Cornell}
V_{\Sigma^+_g}(r) = -\frac{4}{3} \frac{\alpha_s}{r} + \sigma r + V_0 \, ,
\end{equation}
specifically populating the lowest orbitals $\Sigma^+_g(1S,1P,2S,1D)$~\cite{Giron:2019bcs}, a considerable simplification.  Nevertheless, all of the results presented in this paper could be extended to the general case.  In fact, here we restrict to just the states appearing in the orbital $\Sigma^+_g(1S)$, which is believed to accommodate a large fraction of known exotic candidates.  Under this assumption, the overall $J^{PC}$ quantum numbers of the states are entirely determined by combining the spin-0 and spin-1 diquarks:
\begin{eqnarray}
J^{PC} = 0^{++}: & \ & X_0 \equiv \left| 0_\de , 0_\bde \right>_0 \,
, \ \ X_0^\prime \equiv \left| 1_\de , 1_\bde \right>_0 \, ,
\nonumber \\
J^{PC} = 1^{++}: & \ & X_1 \equiv \frac{1}{\sqrt 2} \left( \left|
1_\de , 0_\bde \right>_1 \! + \left| 0_\de , 1_\bde \right>_1 \right)
\, ,
\nonumber \\
J^{PC} = 1^{+-}: & \ & \, Z \  \equiv \frac{1}{\sqrt 2} \left( \left|
1_\de , 0_\bde \right>_1 \! - \left| 0_\de , 1_\bde \right>_1 \right)
\, ,
\nonumber \\
& \ & \, Z^\prime \equiv \left| 1_\de , 1_\bde \right>_1 \, ,
\nonumber \\
J^{PC} = 2^{++}: & \ & X_2 \equiv \left| 1_\de , 1_\bde \right>_2 \,
.
\label{eq:Swavediquark}
\end{eqnarray}
Here, the total $\de(\bde)$ spin is denoted by $s^{\vphantom\dagger}_{\de}(s_{\bde})$, and the overall state total spin $s \! = \! J$ is signified by an outer subscript.  The multiplet $\Sigma^+_g(1S)$ is thus given by 6 states when isospin is ignored, or 12 if both $I \! = \! 0$ and $I \! = \! 1$ are included~\cite{Lebed:2017min}.  The most thorough analysis of the hidden-charm $\Sigma^+_g(1S)$ states in the adiabatic version of the model, which includes $\chi_{c1}(3872)$, $T_{c\bar c 1}(3900)$, and $T_{c\bar c 1}(4020)$ as candidate members, appears in Ref.~\cite{Giron:2021sla}, while other works study negative-parity hidden-charm [$\Sigma^+_g(1P)$] tetraquarks such as $\psi(4230)$~\cite{Giron:2020fvd}, hidden-bottom and $c\bar c s\bar s$ states~\cite{Giron:2020qpb}, $c\bar c c\bar c$ states~\cite{Giron:2020wpx}, hidden-charm, open-strangeness states~\cite{Giron:2021sla}, hidden-charm pentaquarks~\cite{Giron:2021fnl}, and hidden-strangeness states~\cite{Jafarzade:2025qvx}.
  
\subsection{The Diabatic Extension}

Despite extensive work accomplished using the original dynamical diquark model, its founding assumptions assign no important role to the particular di-meson states to which tetraquarks decay (or meson-baryon states to which pentaquarks decay).  And yet, as abundantly illustrated by the discussion surrounding Eq.~(\ref{eq:Xbind}), the masses of many exotic states lie much closer to such thresholds than random chance can easily accommodate.

The diabatic formalism is a rigorous extension of the BO approach to cases in which constituents of the state can interact via distinct potential energy functions that describe different arrangements of the constituents.  It was first developed in the context of atomic and molecular physics (Ref.~\cite{Baer:2006} provides a textbook discussion) as a specific, nonperturbative treatment to incorporate the mixing of coupled-channel contributions from distinct state configurations.  We use a particular implementation of the adiabatic-to-diabatic transformation derived in Ref.~\cite{Baer:2006} and outlined below.  The first use of this diabatic approach in hadronic physics, specifically for the study of heavy-quark exotic states, was implemented by introducing mixing between $Q\bar Q$ states and $(Q\bar q)(\bar Q q)$ di-meson thresholds~\cite{Bruschini:2020voj}.  Note that a study with a similar approach that can be termed diabatic, also using lattice inputs, appeared slightly earlier~\cite{Bicudo:2019ymo}.  The approach was generalized to mixing between $\de\bde$ and di-meson states in Ref.~\cite{Lebed:2022vks}.  Applications to the calculation of mass shifts and decay widths of exotics, and to the effects of the exotic states on the scattering of the di-meson states, have been developed for both underlying fundamental $Q\bar Q$~\cite{Bruschini:2021cty,Bruschini:2021ckr} and $\de\bde$~\cite{Lebed:2024rsi,Lebed:2023kbm} states, respectively.

In each case, the system contains two heavy color sources that interact through light d.o.f., and hence may be described in terms of a nonrelativistic Hamiltonian:
\begin{equation} 
\label{eq:SepHam}
H=K_{\rm heavy} + H_{\rm light} =
\frac{\mathbf{p}^2}{2 \mu_{\rm heavy}} + H_{\rm light}.
\end{equation}
Here, $H_{\rm light}$ contains both the light-field static energy and the heavy-light interaction.  One may seek to separate solutions to the corresponding \schro equation in terms of (gauge-variant) heavy source-pair states  $|\mathbf{r} \rangle$ of separation $\mathbf{r}$ and $H_{ \rm light}$ eigenstates $|\xi_i(\mathbf{r}) \rangle$ (labeled by channel quantum number $i$):
\begin{equation} 
\label{eq:AdExp}
|\psi \rangle = \sum_{i} \int d\mathbf{r} \, \tilde \psi_i
(\mathbf{r}) \, |\mathbf{r} \rangle \:
|\xi_i(\mathbf{r}) \rangle ,
\end{equation}
as suggested by restricting to the BO approximation.  Nevertheless, this expansion remains general because $\{ |\xi_i(\mathbf{r}) \rangle \}$ forms a complete, orthonormal basis for the light d.o.f.\ at any given $\mathbf{r}$, and departures from the assumptions of the BO approximation can be included through configuration mixing: $\langle \xi_j (\mathbf{r}') | \xi_i (\mathbf{r^{\vphantom\prime}}) \rangle \! \neq \! 0$ in general.  Such overlaps become especially important at values of $\mathbf{r}$ near level crossings of potential energy functions.

A convenient yet rigorous generalization of the BO approximation in such circumstances is provided by the \textit{diabatic formalism}~\cite{Baer:2006}.  This method introduces a new expansion for the solution in Eq.~(\ref{eq:AdExp}):
\begin{equation} 
\label{eq:DiaExp}
|\psi \rangle = \sum_{i} \int d\mathbf{r}' \tilde \psi_i
(\mathbf{r}' \! , \mathbf{r}_0) \: |\mathbf{r}' \rangle \:
|\xi_i(\mathbf{r}_0) \rangle,
\end{equation}
where $\mathbf{r}_0$ is a free parameter discussed below.  The basis $\{ |\xi_i(\mathbf{r}) \rangle \}$ remains complete for any choice of $\mathbf{r}$, and in particular for $\mathbf{r}_0$.  Inserting now the expansion Eq.~(\ref{eq:DiaExp}) into the \schro equation and taking inner products with $\langle \xi_j(\mathbf{r}_0) |$, one obtains:
\begin{equation}\label{eq:DiaSchro}
\sum_{i} \left[ - \frac{\hbar^2}{2 \mu_{i}} \de_{ji}  \nabla ^2 +
V_{ji}(\mathbf{r,r_0})-E \de_{ji} \right] \! \tilde \psi_i (\mathbf{r,r_0}) = 0.
\end{equation}

The new ingredient in this \schro equation is the (gauge-variant) coupled-channel potential $V_{ji}$, known as the \textit{diabatic potential matrix}; it is defined by: 
\begin{equation}
V_{ji}(\mathbf{r,r_0}) \equiv \langle \xi_j (\mathbf{r}_0)|
H_{\rm light} (\mathbf{r}) |\xi_i(\mathbf{r}_0) \rangle.
\end{equation}
Since the eigenstates of $H_{\rm light} (\mathbf{r})$ are $|\xi_i(\mathbf{r}) \rangle$ rather than $|\xi_i(\mathbf{r}_0) \rangle$ when $\mathbf{r} \! \neq \! \mathbf{r}_0$, $V_{ji}$ is generally off-diagonal, leading to a system of coupled \schro equations.  One then chooses ${\mathbf{r}}_0$ to lie far from potential-energy level crossings, which allows the states $|\xi_i(\mathbf{r}_0) \rangle$ to be identified with pure, unmixed configurations.  In the {\it diabatic\/} dynamical diquark model, the diagonal elements $V_{ii}$ consist of the static light-field potential energy $V_{\de \bde}(r)$ associated with a pure $\de \bde$ state ({\it i.e.}, the lattice QCD-determined potential $V_{\Lambda^\epsilon_\eta}(r)$ discussed in the previous section) and its corresponding di-meson $(M_1 \overline{M}_2)$ threshold potentials $V_{M_1 \bbar M_2}^{(i)}$, $i = 1, 2, \! \ldots \! , N$.  Explicitly, $V_{ji}$ are collected as 
\begin{equation} \label{eq:FullV}
{\bm{V}}=
\begin{pmatrix}
V_{\de \bde}(\mathbf{r}) & V_{\rm mix}^{(1)}(\mathbf{r})  & \cdots &
V_{\rm mix \vphantom{\bbar M_2}}^{(N)}(\mathbf{r}) \\
V_{\rm mix}^{(1)}(\mathbf{r}) & 
V_{M_1 \bbar M_2}^{(1)}(\mathbf{r}) &
&
\\
\vdots
& & \ddots \\
V_{\rm mix \vphantom{\bbar M_2}}^{(N)}(\mathbf{r}) & & &
V_{M_1 \bbar M_2}^{(N)}(\mathbf{r}) \\
\end{pmatrix}.
\end{equation}
Here, the blank elements represent mixing terms between distinct di-meson configurations, and are taken to be zero.  We set each pure di-meson energy equal to its value for two free, interacting mesons:
\begin{equation} \label{eq:FreeDihadron}
V_{M_1 \bbar M_2}^{(i)}(\mathbf{r}) \to T_{M_1 \bbar M_2} = M_1 + M_2 \, ,
\end{equation}
although one may modify the short-distance interaction $V_{M_1 \bbar M_2}^{(i)}(\mathbf{r})$ between the two mesons in any of the channels to incorporate a mildly attractive potential~\cite{Lebed:2025xbz}, such as might originate through pion-exchange interactions or the effects of triangle singularities.

The off-diagonal elements $V^{(i)}_{\rm mix}(\mathbf{r})$ represent the actual mixing potentials between the pure fundamental ($Q\bar Q$ or $\de\bde$) state and the pure di-meson states.  In the $Q\bar Q$ case, $V^{(i)}_{\rm mix}$ is a color string-breaking potential, which indicates the transition amplitude with which the $Q$ and $\bar Q$ sources with separation $\mathbf{r}$ produce an extra light $q\bar q$ pair from the vacuum, and hence allow a nonzero overlap with the di-meson state.  In the $\de\bde$ case, $V^{(i)}_{\rm mix}$ indicates the transition amplitude for the $\de$ and $\bde$ sources with separation $\mathbf{r}$ to overlap (effectively, a quark rearrangement) with the di-meson state.  In principle, either of these mixing potentials can be determined numerically on the lattice.  However, until such simulations are available, the mixing potentials may be modeled using, {\it e.g.}, Gaussian forms~\cite{Bruschini:2020voj}, which fall off quickly away from thresholds to avoid substantial mixing between distant states:
\begin{equation} \label{eq:Mixpot}
|V_{\rm mix}^{(i)} (r)| = \frac{\Delta}{2}
\exp \! \left\{ -\frac 1 2 \frac{\left[
V^{\vphantom\dagger}_{\de \bde}(r) -
T_{M_1 \bbar M_2 }^{(i)} \right]^2}{(\rho \sigma)^2} \right\} ,
\end{equation}
where $\sigma$ is the string-tension parameter (units of energy squared) that can be extracted from the large-$r$ confining limit of $V_{\de\bde}(r)$ [Eq.~(\ref{eq:Cornell})], $\Delta$ is a parameter (units of energy) indicating the strength of the mixing, and $\rho$ is the (dimensionless) width scale for the level crossing.  Note that this form is purely phenomenological and does not reflect, for example, studies of $Q\bar Q$/heavy-meson-pair mixing potential analyses of lattice data~\cite{TarrusCastella:2022rxb}.  It is also essential to note that, despite pieces of the \schro equation being gauge-variant, the final results for observables such as the overall state mass eigenvalue are gauge-invariant, as they must be.

\section{Threshold Spin Dependence in Heavy-Quark Symmetry} \label{sec:Thresh}

All previous applications of the diabatic dynamical diquark model have employed uniform numerical values for the strength $\Delta$ and width $\rho$ parameters [Eqs.~(\ref{eq:Mixpot})] for all transition potentials between $\de\bde$ and di-meson threshold states.  Of course, one expects somewhat different values of couplings to arise for distinct di-meson pairs.  In the cases of greatest current experimental interest (hidden heavy flavor), the relevant thresholds consist of meson pairs of open heavy-flavor, most notably $D^{(*)} \bar D^{(*)}$ or $B^{(*)} \bar B^{(*)}$.  In the context of heavy-quark spin symmetry (HQSS)~\cite{Manohar_Wise_2023}, open heavy-flavor hadrons ({\it i.e.}, containing a single quark $Q \! = \! c,b$ of mass $m_Q \! \gg \! \Lambda_{\rm QCD}$) differing only by the orientation of spin of $Q$, such as $D$ and $D^*$, are members of the same symmetry multiplet, and are degenerate in mass at leading order [$O(m_Q^1)$], as well as at the first subleading order [$O(m_Q^0 \Lambda_{\rm QCD})$].  The mass splitting $m_{D^*} \! - m_D$, for example, therefore appears at $O(\Lambda_{\rm QCD}^2/m_c)$.  Since $D$ and $D^*$ differ only in the relative spin orientation of their heavy $Q \! = \! c$ quark compared to the light d.o.f., one concludes that HQSS-breaking effects for hadrons with a single heavy quark $Q$ first appear at $O(1/m_Q)$.  Likewise, the kinetic energy of $Q$ in the hadron rest frame is $p^2/2m_Q$, another $O(1/m_Q)$ effect.  Note that these effects correspond to the operators first appearing at $O(1/m_Q)$; in contrast, a great deal of work has gone into developing the full BO effective theory for exotics, including renormalization and matching effects in the $1/m_Q$ expansion.  For recent advances, see Ref.~\cite{Berwein:2024ztx} and references therein.

In contrast, the strict BO approximation assumes stat\-ic, and hence infinitely heavy, sources compared to the light d.o.f\@.  Systematically lifting the BO approximation therefore amounts to consistently incorporating all $1/m_Q$ corrections.  Since HQSS identifies $D , D^*$ (or $B , B^*$) as being two variants of the same state in the heavy-quark limit ({\it i.e.}, the wave function of their light d.o.f., the famous Isgur-Wise function~\cite{Manohar_Wise_2023}, is the same), the effects of distinct thresholds involving mesons within these multiplets are the same up to group-theoretical factors involving which specific states comprise the threshold, and the explicit $O(1/m_Q)$ effects discussed above that enter through HQSS-breaking mass differences and heavy-quark kinetic-energy operators.  The derivation of the formalism that incorporates all of these effects for the diabatic approach to mixing between conventional heavy quarkonium $Q\bar Q$ and $(Q\bar q)(\bar Q q)$ di-meson states is the central purpose of Ref.~\cite{Bruschini:2023zkb}.  Our central purpose in this work is to derive the corresponding formalism for mixing between $\de \bde$ states and the same di-meson states.

Before proceeding, we comment that the explicit results derived in Ref.~\cite{Bruschini:2023zkb} refer only to eigenstates of the lowest BO potential $\Sigma^+_g$, and couplings only to mesons in the lowest HQSS multiplets ($D , D^*$ or $B , B^*$).  As noted in \cite{Bruschini:2023zkb}, however, results for other BO potentials and other HQSS meson (or baryon) multiplets can be treated similarly.  Since, as noted in Sec.~\ref{sec:Model}, all heavy-quark exotic candidates observed to date appear to inhabit the lowest orbitals [$\Sigma^+_g(1S,1P,1D,2S)$] of this BO potential in various flavor sectors~\cite{Giron:2019bcs}, we likewise choose not to incorporate higher HQSS multiplets or BO potentials here.

However, just generalizing the derivation to be applicable to the $\de \bde$ system (accomplished in the next section) still requires four major extensions.  First, the $\de \bde$ system allows a broader range of $J^{PC}$ quantum numbers than $Q\bar Q$, simply by virtue of containing more spin-$\frac 1 2$ quark constituents.  The $\Sigma^+_g(1S)$ multiplet, for example, contains the 6 states listed in Eqs.~(\ref{eq:Swavediquark}): two with $0^{++}$, two with $1^{+-}$, and one each with $1^{++}$ and $2^{++}$.  Second, the light quarks in $\de \bde$ (unlike $Q\bar Q$) allow both $I \! = \! 0$ and $I \! = \! 1$ combinations (and $I \! = \! \frac 1 2$ for open-strangeness states).  That isospin dependence cannot be neglected for exotic states is highlighted by the fact that the famous $\chi_{c1}(3872)$ appears to be pure $I \! = \! 0$, since experiment has failed to detect any nearby charged (hence $I \! = \! 1$) partners~\cite{Aubert:2004zr,Choi:2011fc}, while the nearby $T_{c\bar c 1}(3900)$, $T_{c\bar c 1}(4020)$ multiplets have charged members, and hence are $I=1$.  Third, obtaining the correct degree of mixing between $\de\bde$ [$(Qq)(\bar Q \bar q)$] and di-meson [$(Q\bar q)(\bar Q q)$] states, in contrast with that of conventional-quarkonium [$(Q\bar Q)$] states, requires the derivation of variants of Fierz reordering identities that are much less well-known than the standard textbook form~\cite{Peskin:1995ev}.  The fourth distinction, first noted in Ref.~\cite{Lebed:2022vks} and discussed in Sec.~\ref{sec:Model}, is dynamical: The mixing of di-meson states with $Q\bar Q$ results from string breaking, while their mixing with $\de \bde$ results from quark rearrangement.  And of course, the heavy sources in the $\de\bde$ case are diquark quasiparticles rather than individual heavy quarks.

In the following subsections we parallel the derivation of Ref.~\cite{Bruschini:2023zkb}, generalizing its approach where necessary in order to be applicable to the $\de\bde$ system.

\subsection{BO Configurations and Quantum Numbers}

First, in order to accommodate specific states of the $Q\bar Q$ pair, Ref.~\cite{Bruschini:2023zkb} generalizes the conventional $\Lambda^\epsilon_\eta$ notation by including $Q\bar Q$ spin and $CP$ eigenvalues.  The $Q\bar Q$ spin ${\bf S}_{Q\bar Q}$ and the light-d.o.f.\ angular momentum ${\bf J}_{\rm light}$ are combined:
\begin{equation} \label{eq:TotSpin}
{\bf S} \equiv {\bf S}_{Q\bar Q} + {\bf J}_{\rm light} ,
\end{equation}
with the Casimir eigenvalue of ${\bf S}$ being labeled as $s$, and the eigenvalue of its projection along the source axis $\hat{\bm{r}}$ being defined through:
\begin{equation} \label{eq:lambdadef}
\lambda \equiv {\bf S} \cdot \hat{\bm{r}} .
\end{equation}
Note that, unlike $\Lambda$, the quantum number $\lambda$ includes sign information.  In the case that the light d.o.f.\ are assumed to be in the trivial ($\Sigma^+_g$) BO potential where ${\bf J}_{\rm light} \! = \! 0$, then $\lambda$ is simply the axial spin projection of the heavy-quark spin ${\bf S}_{Q\bar Q}$, according to Eq.~(\ref{eq:TotSpin}).

Similarly, the $CP$ eigenvalue for all conventional $Q\bar Q$ states is given by $(-1)^{s+1}$, and again assuming that the light d.o.f.\ are in the $\Sigma^+_g$ ($CP \! = \! +$) configuration, the total $CP$ eigenvalue $\eta$ for $Q\bar Q$ is given by $(-1)^{s+1}$.  Lastly, Ref.~\cite{Bruschini:2023zkb} shows that the total eigenvalue $\epsilon$ for heavy sources of known spin and intrinsic parity combined with a specified light-d.o.f.\ configuration is completely fixed and hence redundant.  Thus, the new notation for describing the BO configuration is given by $\lambda_\eta$.

The quantum numbers for the di-meson states are exactly as in Ref.~\cite{Bruschini:2023zkb}: $|\lambda| \! \leq \! s$, the intrinsic $CP$ eigenvalue $\eta$ for $D\bar D$ or $D^* \! \bar D^*$ in a state of total spin $s$ is given by $(-1)^s$, and $D \bar D^*$ states can have either $CP \! = \! \pm$.  These results hold for di-meson states whether they couple to $Q\bar Q$, $\de\bde$, or both.

Using the $\eta \! = \! CP$ and $s$ eigenvalues of the $\de\bde$ states listed in Eqs.~(\ref{eq:Swavediquark}), the constraint $|\lambda| \! \leq \! s$, and the assumption of the light d.o.f.\ in the ground state $\Sigma^+_g$, one obtains the complete set of $\lambda_\eta$ eigenvalues for each $\de\bde$ state as well.  Given $\lambda_\eta$, we obtain the allowed values of $s$ for each $\de\bde$ and $D^{(*)}\bar D^{(*)}$ state, as presented in Table~\ref{tab:Configs}.

To obtain the allowed total $J^{PC}$ value for the full physical state in the $\ell^{\rm th}$ partial wave, one needs only to combine the total system spin $s$ with the orbital angular momentum $\ell$ of the heavy sources (since the orbital angular momentum of the light d.o.f.\ is already included  in ${\bf J}_{\rm light}$) in order to obtain total $J$, and to compute $P \! = \! (-1)^\ell$ and $C \! = \! \eta \, (-1)^\ell$ for either the $\de\bde$ or the di-meson state.  In general, the partial wave $\ell$ for the fundamental $\de\bde$ system and the partial wave $\ell^\prime$ for the di-meson system may be expected to differ, but parity conservation requires this difference to be an even integer.\footnote{In the $Q\bar Q$ case, $P \! = \! (-1)^{\ell + 1}$, so that parity conservation requires $\ell$ and $\ell^\prime$ to differ by an odd integer.}  All partial waves for $J \! = \! 0, 1, 2$ states obtained in this way are examined in Sec.~\ref{sec:ExplicitVs}.

\begin{table}[h]
\centering
\renewcommand{\arraystretch}{1.5}
\caption{Allowed total spin eigenvalues $s$ of $\de\bde$ sources [including corresponding state labels from Eqs.~(\ref{eq:Swavediquark})] and of di-meson sources for the given BO quantum numbers $\lambda_\eta$ defined in the text.  In the $\de\bde$ case, a light-QCD configuration $\Sigma_g^+$ combines with the spins of the heavy sources to produce the specified $\lambda_\eta$ eigenvalues.\label{tab:Configs}}
\begin{tabular*}{\columnwidth}{@{\extracolsep{\stretch{1}}}cccccc}
\hline\hline
Source			 & $0_g$	 & $\pm1_g$	& $\pm2_g$	& $0_u$	& $\pm1_u$	\\
\hline
$\de\bde$		 & 0,1,2   &  1,2		& 2  		& 1		& 1  		\\
& $X^{(\prime)}_0 \! , X_{1,2}$ & $X_{1,2}$   & $X_2$        & $Z^{(\prime)}$
& $Z^{(\prime)}$ \\
$D \bar{D}$		 & 0		 & --- 		& ---		& ---	& --- 		\\
$D \bar{D}^*$        & 1	 & 1			& ---		& 1		& 1			\\
$D^* \! \bar{D}^*$ & 0,2     & 2			& 2			& 1		& 1			\\
\hline\hline
\end{tabular*}
\end{table}

\subsection{Overlap Operators and Fierz Reordering}

The next step in the derivation of the formalism is to determine the overlap between the fundamental (here, $\de\bde$) operators with $D^{(*)} \bar D^{(*)}$ operators of the same $\lambda_\eta$ quantum numbers, as listed in Table~\ref{tab:Configs}.  In the case of $\bar Q \Gamma Q$ operators~\cite{Bruschini:2023zkb} ($\Gamma$ denoting specific Dirac structures, and to be formally accurate, also including a Wilson line to connect the heavy $Q\bar Q$ sources with separation {\bf r} in order to maintain gauge invariance), the di-meson interpolating operators of form $(\bar q \, \Gamma_1 Q)(\bar Q \, \Gamma_2 q)$ undergo a Fierz reordering into all-heavy/all-light bilinear combinations of the form $(\bar Q \, \Gamma^\prime_1 Q)(\bar q \, \Gamma^\prime_2 q)$.  The beginning of Appendix~\ref{sec:Fierz} establishes conventions for the Dirac matrices used in this work, and presents the fundamental Fierz identity upon which all others are based in Eq.~(\ref{eq:ProtoFierz}).

The terms for which $\bar q \, \Gamma^\prime_2 q$ has the same quantum numbers as the vacuum (and $\Gamma_1^{\prime} \! \to \!\Gamma$) provide the path for overlap with the original $\bar Q \Gamma Q$ operator.  Its overlap strength is given by the coefficient obtained through the Fierz reordering, which in the $Q\bar Q$ case is the familiar form given in Eq.~(\ref{eq:Fierz0}).  To be precise, the static positive-energy heavy-quark field is obtained from the initial field $Q$ via the projection operator $P_+$, where
\begin{equation} \label{eq:EProj}
P_\pm \equiv \frac{1 \pm \gamma^0}{2} .
\end{equation} 
In the heavy-quark limit, $Q \! = \! P_+ Q$ and $\bar Q \! = \! \bar Q P_-$.  The bilinear $\bar q \, \Gamma^\prime_2 q$ that overlaps with the vacuum is~\cite{Bruschini:2023zkb}:
\begin{equation} \label{eq:VacOp}
\bar q P_- \, \hat{\bm{r}} \cdot \! {\bm{\gamma}} \, q .
\end{equation}
The overlap coefficients with Eq.~(\ref{eq:VacOp}) are compiled in Eqs.~(4)-(10) of Ref.~\cite{Bruschini:2023zkb}, and form the mixing (off-diag\-onal) elements of the string-breaking amplitude matrices $\bm{G}^{\eta, \lambda}(r)$ given in Eqs.~(13)-(17) of Ref.~\cite{Bruschini:2023zkb}.  The diagonal elements of $\bm{G}^{\eta, \lambda}(r)$ consist of the adiabatic potential $V_{Q\bar Q}(r)$ and the (trivial) di-hadron potentials of Eq.~(\ref{eq:FreeDihadron}).  More precisely, the zero of energy is chosen to be twice the spin-averaged mass $m$ of the ground-state heavy-quark multiplet,
\begin{equation} \label{eq:MassAve}
m \equiv \frac 1 4 \left( m_D + 3 m_{D^*} \! \right) .
\end{equation}
Upon taking
\begin{equation} \label{eq:MassSplit}
\Delta \equiv m_{D^*} \! - m_D ,
\end{equation}
then the free di-hadron potential elements are simply $-\frac 3 2 \Delta, -\frac 1 2 \Delta, +\frac 1 2 \Delta$ for $D\bar D$, $D \bar D^*$, $D^* \! \bar D^* \!$, respectively.

The analogous calculation with $\de\bde$ as the fundamental state is more complicated.  It requires the Fierz reordering of $\de\bde \! = \! (q \, Q)(\bar Q \, \bar q^\prime)$ [the prime on $q^\prime$ allowing for including isospin dependence, and again with the presence of Wilson-line operators to maintain gauge invariance being understood], where parentheses indicate color-$\bar{\bf 3}$({\bf 3}) combinations, respectively.  It is convenient to introduce charge-conjugate fermion fields $\psi^c$ and $\bar \psi^c$ in the usual manner (Appendix~\ref{sec:ChargeConj} summarizes their key properties).  The full operators corresponding to the normalized diquark quasiparticles of interest, exhibiting explicit color indices and $J^P$ quantum numbers, are
\begin{eqnarray}
\frac{1}{\sqrt{2}} \, \epsilon^{ijk} {\bar q}^{\, c}_j \, i\gamma_5 \, Q_k , & & \ \ \ \
\frac{1}{\sqrt{2}} \, \epsilon^{imn} {\bar Q}_m \, i\gamma_5 \, q^{\prime \, c}_n \ \ \ (0^+) , \nonumber \\
\frac{1}{\sqrt{2}} \, \epsilon^{ijk} {\bar q}^{\, c}_j \gamma^\mu Q_k , & & \ \ \ \
\frac{1}{\sqrt{2}} \, \epsilon^{imn} {\bar Q}_m \gamma^\mu q^{\prime \, c}_n \ \ \ \ \, (1^+) .
\label{eq:DiquarkOps}
\end{eqnarray}

Since each diquark contains a heavy quark of flavor $Q$, the mass difference between the corresponding quasiparticles is expected to scale as $1/m_Q$, thus substantially suppressing the distinction between ``good'' and ``bad'' diquarks~\cite{Jaffe:2004ph} relevant to the light-quark sector.  If the two quark flavors in either of the diquarks are identical (as is the case for the $(cc)$ diquark in the corresponding interpretation of the double-charmed state $T_{cc}^+$~\cite{Lebed:2024zrp}), then the $\frac{1}{\sqrt{2}}$ normalization factor is replaced with $\frac{1}{2}$.\footnote{These factors were incorrectly omitted in Ref.~\cite{Lebed:2024zrp}.}

The full $\de\bde$ operators appearing in the diabatic potential matrix $V_{ji}$ should be compared to those appearing in the initial application of Ref.~\cite{Bruschini:2023zkb}, which represent mixing between $Q\bar Q$ operators and di-meson states.  Most significantly, the color structure of the former ($\bar Q_i \, Q_i$) is simpler than that induced in the $\de\bde$ operators formed from products of the $\de$ and $\bde$ structures given in Eqs.~(\ref{eq:DiquarkOps}).  To develop the $\de\bde$ diabatic potential matrix, one first uses the identity
\begin{equation}
\epsilon^{ijk} \epsilon^{imn} = \delta^{jm} \delta^{kn} - \delta^{jn} \delta^{km} ,
\end{equation}
to turn $\de\bde$ operators into combinations of bilinear operators of the forms $(\bar Q_i Q_i)(\bar q^\prime_j q_j)$ and $(\bar Q_{i \,} q_i)(\bar q^\prime_j Q_j)$, suppressing Dirac structures that appear in these bilinears.  The first of these combinations has the structure of heavy-quarkonium/light-meson pairs, and in principle could be used to examine couplings of tetraquark states to such meson pairs.  However, heavy quarkonium is expected to be rather more compact than open heavy-flavor mesons (and certainly than light-quark mesons), so that one expects $(\bar Q_i Q_i)(\bar q^\prime_j q_j)$, particularly the large state $(\bar q^\prime_j q_j)$ state, to have a much smaller wave-function overlap with the full tetraquark state, where the heavy- and light-quark pairs have comparable separation.

We therefore treat the $(\bar Q_i Q_i)(\bar q^\prime_j \, q_j)$ operators as requiring an additional Fierz reordering into operators of the form $(\bar Q_i \, q_i)(\bar q^\prime_j Q_j)$.  This reordering is nontrivial in two distinct ways: First, rearranging the (suppressed) Dirac structures requires the derivation of new types of Fierz reorderings, in which the products of the two bilinears do not transform as Lorentz scalars; the necessary additional forms are presented in Eqs.~(\ref{eq:Fierz1}) and (\ref{eq:Fierz2}).  Second, the color indices in reordering the operators $(\bar Q_i \, Q_i)(\bar q^\prime_j q_j)$ do not coincide with those of $(\bar Q_i \, q_i)(\bar q^\prime_j Q_j)$, and thus the SU(3)$_{\rm color}$ Fierz reordering identity ({\it i.e.}, the color-generator $\lambda^a$ completeness relation) Eq.~(\ref{eq:ColorFierz}) must also be applied.

These lengthy manipulations produce:
%
\begin{widetext}
\begin{eqnarray}
\lefteqn{ \left( \epsilon^{ijk} \bar q^c_j \, i \gamma_5 \, Q_k \right) \left( \epsilon^{imn} \bar Q_m \, i \gamma_5 \, q^{\prime \, c}_n \right)} & & \nonumber \\
& = & +\frac{1}{6} \left[ + \left( \bar q^\prime Q \right) \! \left( \bar Q q \right) - \left( \bar q^\prime i \gamma_5 Q \right) \! \left( \bar Q \, i \gamma_5 q \right) + \left( \bar q^\prime \gamma^\mu Q \right) \! \left( \bar Q \gamma_\mu q \right) + \left( \bar q^\prime \gamma^\mu \gamma_5 Q \right) \! \left( \bar Q \gamma_\mu \gamma_5 q \right) -\frac 1 2 \left( \bar q^\prime \sigma^{\mu\nu} Q \right) \! \left( \bar Q \sigma_{\mu\nu} q \right) \right] \nonumber \\
& & -\frac{1}{8} \big[ \mbox{same expression as in the previous brackets, with $\lambda^a$ inserted into each bilinear} \big] ,
\label{eq:0+0+}
\end{eqnarray}
\end{widetext}
%
\begin{widetext}
\begin{eqnarray}
\lefteqn{ \left( \epsilon^{ijk} \bar q^c_j \, \gamma_\mu Q_k \right) \left( \epsilon^{imn} \bar Q_m \, i \gamma_5 \, q^{\prime \, c}_n \right)} & & \nonumber \\
& = & +\frac{1}{6} \Big[ -i \left( \bar q^\prime Q \right) \! \left( \bar Q \gamma_\mu \gamma_5 \, q \right) + \left( \bar q^\prime \sigma_{\mu\nu} Q \right) \! \left( \bar Q \, \gamma^\nu \gamma_5 q \right) + \left( \bar q^\prime i \gamma_5 Q \right) \! \left( \bar Q \gamma_\mu q \right) +i \left( \bar q^\prime i \sigma_{\mu\nu} \gamma_5 Q \right) \! \left( \bar Q \gamma^\nu q \right) \nonumber \\
& & \hspace{2.2em} + \, i \left( \bar q^\prime \gamma_\mu \! \gamma_5 \, Q \right) \! \left( \bar Q q \right) - \left( \bar q^\prime \gamma^\nu \gamma_5 \, Q \right) \! \left( \bar Q \sigma_{\mu\nu} q \right) - \left( \bar q^\prime \gamma_\mu Q \right) \! \left( \bar Q \, i \gamma_5 \, q \right) -i \left( \bar q^\prime \gamma^\nu Q \right) \! \left( \bar Q \, i \sigma_{\mu\nu} \gamma_5 q \right) \Big] \nonumber \\
& & +\frac{1}{8} \big[ \mbox{same expression as in the previous brackets, with $\lambda^a$ inserted into each bilinear} \big] ,
\label{eq:1+0+}
\end{eqnarray}
\end{widetext}
%
\begin{widetext}
\begin{eqnarray}
\lefteqn{ \left( \epsilon^{ijk} \bar q^c_j \, i \gamma_5 \, Q_k \right) \left( \epsilon^{imn} \bar Q_m \gamma_\mu \, q^{\prime \, c}_n \right)} & & \nonumber \\
& = & +\frac{1}{6} \Big[ -i \left( \bar q^\prime Q \right) \! \left( \bar Q \gamma_\mu \gamma_5 \, q \right) - \left( \bar q^\prime \sigma_{\mu\nu} Q \right) \! \left( \bar Q \, \gamma^\nu \gamma_5 q \right) - \left( \bar q^\prime i \gamma_5 Q \right) \! \left( \bar Q \gamma_\mu q \right) +i \left( \bar q^\prime i \sigma_{\mu\nu} \gamma_5 Q \right) \! \left( \bar Q \gamma^\nu q \right) \nonumber \\
& & \hspace{2.2em} + \, i \left( \bar q^\prime \gamma_\mu \gamma_5 \, Q \right) \! \left( \bar Q q \right) + \left( \bar q^\prime \gamma^\nu \gamma_5 \, Q \right) \! \left( \bar Q \sigma_{\mu\nu} q \right) + \left( \bar q^\prime \gamma_\mu Q \right) \! \left( \bar Q \, i \gamma_5 \, q \right) -i \left( \bar q^\prime \gamma^\nu Q \right) \! \left( \bar Q \, i \sigma_{\mu\nu} \gamma_5 q \right) \Big] \nonumber \\
& & -\frac{1}{8} \big[ \mbox{same expression as in the previous brackets, with $\lambda^a$ inserted into each bilinear} \big] ,
\label{eq:0+1+}
\end{eqnarray}
\end{widetext}
%
\begin{widetext}
\begin{eqnarray}
\lefteqn{ \left( \epsilon^{ijk} \bar q^c_j \, \gamma_\mu Q_k \right) \left( \epsilon^{imn} \bar Q_m \gamma_\nu \, q^{\prime \, c}_n \right)} & & \nonumber \\
& = & +\frac{1}{6} \Big[
-g_{\mu\nu} \! \left( \bar q^\prime Q \right) \! \left( \bar Q q \right) +g_{\mu\nu} \! \left( \bar q^\prime i \gamma_5 \, Q \right) \! \left( \bar Q \, i \gamma_5 \, q \right) \nonumber \\
& &  \hspace{2.1em} -\left( \bar q^\prime \gamma_\mu Q \right) \! \left( \bar Q \gamma_\nu q \right) - \left( \bar q^\prime \gamma_\nu \, Q \right) \! \left( \bar Q \gamma_\mu q \right) + g_{\mu\nu} \! \left( \bar q^\prime \gamma^\rho Q \right) \! \left( \bar Q \, \gamma_\rho \, q \right) \nonumber \\
& &  \hspace{2.1em} +\left( \bar q^\prime \gamma_\mu \gamma_5 \, Q \right) \! \left( \bar Q \gamma_\nu \gamma_5 \, q \right) + \left( \bar q^\prime \gamma_\nu \gamma_5 \, Q \right) \! \left( \bar Q \gamma_\mu \gamma_5 \, q \right) - g_{\mu\nu} \! \left( \bar q^\prime \gamma^\rho \gamma_5 \, Q \right) \! \left( \bar Q \, \gamma_\rho \gamma_5 \, q \right) \nonumber \\
& & \hspace{2.1em} + \, i \left( \bar q^\prime Q \right) \! \left( \bar Q \sigma_{\mu\nu} q \right) +i \left( \bar q^\prime i \gamma_5 \, Q \right) \! \left( \bar Q \, i \sigma_{\mu\nu} \gamma_5 \,  q \right) -i \epsilon_{\mu\nu\rho\sigma} \! \left( \bar q^\prime \gamma^\rho Q \right) \! \left( \bar Q \gamma^\sigma \gamma_5 \, q \right) \nonumber \\
& & \hspace{2.1em} + \, i \left( \bar q^\prime \sigma_{\mu\nu} Q \right) \! \left( \bar Q q \right) +i \left( \bar q^\prime i \sigma_{\mu\nu} \gamma_5 \, Q \right) \! \left( \bar Q i \gamma_5 \, q \right) +i \epsilon_{\mu\nu\rho\sigma} \! \left( \bar q^\prime \gamma^\rho \gamma_5 Q \right) \! \left( \bar Q \gamma^\sigma q \right) \nonumber \\
& & \hspace{2.1em} - \frac{1}{2} \, g_{\mu\nu} \! \left( \bar q^\prime \sigma^{\rho\sigma} Q \right) \! \left( \bar Q \,\sigma_{\rho\sigma} q \right) + \left( \bar q^\prime \sigma_{\mu\rho} \, Q \right) \! \left( \bar Q \, \sigma_\nu{}^\rho q \right) + \left( \bar q^\prime \sigma_{\nu\rho} \, Q \right) \! \left( \bar Q \, \sigma_\mu{}^\rho q \right) \Big] \nonumber \\
& & -\frac{1}{8} \big[ \mbox{same expression as in the previous brackets, with $\lambda^a$ inserted into each bilinear} \big] .
\label{eq:1+1+}
\end{eqnarray}
\end{widetext}
The relations Eqs.~(\ref{eq:0+0+})-(\ref{eq:1+1+}) provide the most important intermediate results for obtaining the diabatic potential matrices that describe mixing between $\de\bde$ and di-meson states in various spin sectors.  One simply forms linear combinations of Eqs.~(\ref{eq:0+0+})-(\ref{eq:1+1+}) corresponding to the combinations in Eqs.~(\ref{eq:Swavediquark}) that produce tetraquark states of given $J^{PC}$ quantum numbers, a task we perform in the next subsection.  The bilinear combinations on the right-hand side then indicate the degree of overlap with di-meson states corresponding to the same operator structure.\footnote{The terms with explicit $\lambda^a$ factors naively suggest coupling to hybrid-meson pairs, but can also be interpreted as operators that allow long-distance strong (color-nonsinglet) rescattering into conventional meson pairs.}

The universal amplitude $h(r)$ for rearrangement of the $\de\bde$ state of separation $r$ to a di-meson state serves as this model's analogue to the Ref.~\cite{Bruschini:2023zkb}'s universal function $g(r)$ describing the string-breaking amplitude between $Q\bar Q$ and di-meson states after $Q\bar Q$ spin quantum numbers are factored out, so that $g(r)$ is a type of reduced matrix element in the Wigner-Eckhart sense (See Ref.~\cite{Bruschini:2023zkb}, Eq.~(3.2) and the subsequent text).  In particular, the calculation of either one requires the computation of a Wilson line connecting the two heavy sources (in the present case, $\de\bde$), which is shifted by the Fierz reordering into the correlator that defines $h(r)$ or $g(r)$, respectively, and can presumably be carried out in lattice QCD\@.  In addition, the approach to the heavy-quark limit is implemented slightly differently for the operators on the two sides Eqs.~(\ref{eq:0+0+})-(\ref{eq:1+1+}), since the masses $m_{\de,\bde}$ differ from $m_Q$ by an $O(m_Q^0)$ amount supplied by the light-quark mass and the diquark binding energy.  In a complete and rigorous treatment, such effects propagate into the right-hand sides of Eqs.~(\ref{eq:0+0+})-(\ref{eq:1+1+}) and ultimately contribute to the function $h(r)$; their leading-order in $m_Q$ contribution is to introduce heavy-quark projection operators [Eqs.~(\ref{eq:EProj})]: $Q \! \to \! P_+ Q$ and $\bar Q \! \to \bar Q P_-$.  However, since our goal is to develop expressions from which overlaps with di-meson states are simplest to extract, we do not exhibit these complications explicitly in Eqs.~(\ref{eq:0+0+})-(\ref{eq:1+1+}).

Furthermore, since the diabatic potential is a generalization of the ${\bf 3}$-$\bar{\bf 3}$ potential $V_{Q\bar Q}(r)$ [{\it e.g.}, Eq.~(\ref{eq:Cornell})] and already contains the relevant color-dependent factors of the fundamental interaction, the $\de$ and $\bde$ operators should not be normalized with respect to their free color index $i$ in the same way as if they were one-particle states.

\subsection{Interaction Potential Mixing Matrices}
\label{subsec:MixingMatrices}

For each given set of $\eta, \lambda$ eigenvalues, Ref.~\cite{Bruschini:2023zkb} constructs a string-breaking amplitude matrix $\bm{G}^{\eta, \lambda}$ such that its diagonal elements consist of the fundamental potential $V_{\de\bde}(r)$ and the di-meson masses relative to the spin-multiplet average, as discussed in the previous subsection.  The off-diagonal elements of $\bm{G}^{\eta, \lambda}$ in our case are obtained from Eqs.~(\ref{eq:0+0+})-(\ref{eq:1+1+}), the first step being to identify the particular combinations corresponding to the $J^{PC}$ quantum numbers of the $\de\bde$ states in Eqs.~(\ref{eq:Swavediquark}).

Specifically, we begin by identifying the spherical-tensor components $p$ of the vector (rank-1 tensor) operators with conventional normalization~\cite{Edmonds:1957}, as applied to the Dirac $\gamma$ matrices.  The $p \! = \! \pm 1$ components are obtained through the combinations:
\begin{equation} \label{eq:gamma+-}
\gamma^\pm = \mp \frac{1}{\sqrt{2}} \left( \gamma^1 \pm i \gamma^2 \right) \, ,
\end{equation}
and $\gamma^3$ provides the properly normalized $p \! = \! 0$ component.  Explicitly, using the $1^+$ ($\gamma^p$) entries of Eqs.~(\ref{eq:DiquarkOps}), we define the rank-1 operators:
\begin{equation}
U^p \equiv \frac{1}{\sqrt{2}} \, \epsilon^{ijk} {\bar q}^{\, c}_j \gamma^p Q_k , \ \
V^p \equiv \frac{1}{\sqrt{2}} \, \epsilon^{imn} {\bar Q}_m \gamma^p q^{\prime \, c}_n ,
\end{equation}
while the $0^+$ ($i\gamma_5$) entries of Eqs.~(\ref{eq:DiquarkOps}) are rank-0.

This information is already sufficient to extract off-di\-agonal elements of $\bm{G}^{\eta, \lambda}$ for the cases corresponding to the combinations $X_0$, $X_1$, and $Z$ in Eqs.~(\ref{eq:Swavediquark}), since each of these states contains a diquark with spin 0.  However, one must take care to note that the relative signs between the kets in the definitions of $X_1$ and $Z$ in Eqs.~(\ref{eq:Swavediquark}) indicate that the corresponding states are assumed to be $CP$ conjugates of each other, an issue that does not arise for any of the other states in Eqs.~(\ref{eq:Swavediquark}).  On the other hand, the spatial ($\mu \! \to \! p$) components of the explicit diquark operators in Eqs.~(\ref{eq:1+0+})-(\ref{eq:0+1+}) are $CP$ conjugates of each other but with a relative sign difference, since vector bilinears have $C \! = \! -$ and their spatial components have $P=-$, while pseudoscalar bilinears have $C \! = \! +$ and $P \! = \! -$~\cite{Peskin:1995ev}.  Thus, di-meson overlaps obtained from Eq.~(\ref{eq:0+1+}) with the states $X_1$ and $Z$  must be multiplied by an additional factor of $-1$ compared to those obtained from Eq.~(\ref{eq:1+0+}).

For the remaining combinations $X^\prime_0$, $Z^\prime$, and $X_2$, one needs only the decomposition of the product of two rank-1 tensors with components $U^{p_1}$ and $V^{p_2}$ into rank-$k$ tensors $T^{(k)}$, $k=0$, $1$, and $2$, respectively, with components $p$.  The general expression is~\cite{Edmonds:1957}:
\begin{equation}
T^{(k), \, p} = \sum_{m_1, m_2} \left( \! \begin{array}{cc|@{\hspace{0.4em}}c} 1 & 1 & k \\ p_1 & p_2 & p \end{array} \! \right) U^{p_1} V^{p_2} \, ,
\end{equation}
where the object in parentheses denotes a conventional Clebsch-Gordan coefficient.  Explicitly,
\begin{eqnarray}
T^{(0),0}        & = & +\frac{1}{\sqrt{3}} \left( U^+ V^- \! - U^0 V^0 \! + U^- V^- \right) \, , \nonumber \\
T^{(1),\pm 1} & = & \pm \frac{1}{\sqrt{2}} \left( U^\pm V^0 \! - U^0 V^\pm \right) \, ,
\nonumber \\
T^{(1),0}         & = & +\frac{1}{\sqrt{2}} \left( U^+ V^- \! - U^- V^+ \right) \, ,
\nonumber \\
T^{(2),\pm 2} & = & +U^\pm V^\pm \, ,
\nonumber \\
T^{(2),\pm 1} & = & +\frac{1}{\sqrt{2}} \left( U^\pm V^0 \! + U^0 V^\pm \right) \, , \nonumber \\
T^{(2)_,0}       & = & +\frac{1}{\sqrt{6}} \left( U^+ V^- \! + 2 \, U^0 V^0 \! + U^- V^+ \right) \, .
\label{eq:TensorDecomp}
\end{eqnarray}

The di-meson interpolating operators assume essentially the same forms as those introduced in Ref.~\cite{Bruschini:2023zkb}, except that we use the labels ${\cal D}$ to refer to the charm system.  Indicating the $\eta$ eigenvalue as a superscript (when necessary), the $s,\lambda$ eigenvalues as subscripts, and the number of $D^*$ mesons in the di-meson state via the number of $*$ superscripts, one has:
\begin{eqnarray}
{\cal D}_{0,0} & = & \left( \bar Q \, i\gamma_5 q \right) \! \left( \bar q \, i \gamma_5 Q \right) , \nonumber \\
{\cal D}^{*g}_{1,\lambda} & = & \frac{1}{\sqrt{2}} \left[ \left( \bar Q \gamma^\lambda q \right) \! \left( \bar q \, i \gamma_5 Q \right) - \left( \bar q \, i \gamma_5 Q \right) \! \left( \bar Q \gamma^\lambda q \right) \right] \! , \nonumber \\
{\cal D}^{*u}_{1,\lambda} & = & \frac{1}{\sqrt{2}} \left[ \left( \bar Q \gamma^\lambda q \right) \! \left( \bar q \, i \gamma_5 Q \right) + \left( \bar q \, i \gamma_5 Q \right) \! \left( \bar Q \gamma^\lambda q \right) \right]  \! ,
\label{eq:DiMeson1}
\end{eqnarray}
for each of $\lambda \! \in \! \left\{ +, -, {0} \right\}$, where the relative signs follow the same pattern as described above to form specific $CP$ eigenstates; and using Eqs.~(\ref{eq:TensorDecomp}), one has:
\begin{widetext}
\begin{eqnarray}
{\cal D}^{**}_{0,0} & = & \frac{1}{\sqrt{3}} \left[ \left( \bar Q \gamma^+ q \right) \! \left( \bar q \gamma^- Q \right) - \left( \bar q \, \gamma^3 Q \right) \! \left( \bar Q \gamma^3 q \right) + \left( \bar Q \gamma^- q \right) \! \left( \bar q \gamma^+ Q \right) \right] , \nonumber \\
{\cal D}^{**}_{1,0} & = & \frac{1}{\sqrt{2}} \left[ \left( \bar Q \gamma^+ q \right) \! \left( \bar q \gamma^- Q \right) - \left( \bar Q \gamma^- q \right) \! \left( \bar q \gamma^+ Q \right) \right] , \nonumber \\
{\cal D}^{**}_{1,\pm 1} & = & \pm \frac{1}{\sqrt{2}} \left[ \left( \bar Q \gamma^\pm q \right) \! \left( \bar q \gamma^3 Q \right) - \left( \bar Q \gamma^3 q \right) \! \left( \bar q \gamma^\pm Q \right) \right] , \nonumber \\
{\cal D}^{**}_{2,0} & = & \frac{1}{\sqrt{6}} \left[ \left( \bar Q \gamma^+ q \right) \! \left( \bar q \gamma^- Q \right) +2 \left( \bar q \, \gamma^3 Q \right) \! \left( \bar Q \gamma^3 q \right) + \left( \bar Q \gamma^- q \right) \! \left( \bar q \gamma^+ Q \right) \right] , \nonumber \\
{\cal D}^{**}_{2,\pm 1} & = & \frac{1}{\sqrt{2}} \left[ \left( \bar Q \gamma^\pm q \right) \! \left( \bar q \gamma^3 Q \right) + \left( \bar Q \gamma^3 q \right) \! \left( \bar q \gamma^\pm Q \right) \right] , \nonumber \\
{\cal D}^{**}_{2,\pm 2} & = & \left( \bar Q \gamma^\pm q \right) \! \left( \bar q \gamma^\pm Q \right) .
\label{eq:DiMeson2}
\end{eqnarray}
\end{widetext}
These combinations correspond one-to-one with the entries for $D^{(*)} \bar D^{(*)}$ states in Table~\ref{tab:Configs}.

To obtain the numerical overlap coefficients with specific di-meson states in Eqs.~(\ref{eq:DiMeson1})-(\ref{eq:DiMeson2}) from Eqs.~(\ref{eq:0+0+})-(\ref{eq:1+1+}), one first notes structures that obviously serve as interpolating operators for mesons of known $J^P$: For example, $\bar Q \gamma^+ \! q$ clearly serves as an interpolating operator for a $J^P \! = \! 1^-$ ($Q\bar q$) meson with spin polarization $+1$.  However, several of the other bilinear structures less obviously lead to overlaps with $0^-$ and $1^-$ states, most notably in the heavy-quark limit.  As obtained in Appendix~\ref{subsec:DiracWeyl} [Eqs.~(\ref{eq:EquivP+})-(\ref{eq:EquivP-})], various Dirac structures form equivalence classes when acting to the right upon $Q \! = \! P_+ Q$ or to the left upon $\bar Q \! = \! \bar Q P_-$.   Thus, since $\gamma^+ \! \equiv \! +i \sigma^{0+} \! \equiv \! +i \sigma^{+3} \gamma_5$ when acting upon $P_+$, then an interpolating operator for a $J^P \! = \! 1^-$ ($Q\bar q$) meson with spin polarization $+1$ is provided not only by $\bar Q \gamma^+ \! q$, but also by $\bar Q \, \sigma^{0+} \! q$ and $\bar Q \, \sigma^{+3} \gamma_5 q$.\footnote{Such a possibility in the diabatic formalism was first noted in Ref.~\cite{Lebed:2024zrp}.}

Using these equivalences and restricting to ground-state ($0^-, 1^-$) di-meson pairs (as noted in the beginning of this section), one obtains the following overlap results for the $\de\bde$ states defined in Eqs.~(\ref{eq:Swavediquark}):
\begin{eqnarray}
X_0              & : \ & -\frac 1 6 {\cal D}_{0,0} +\frac{1}{2\sqrt{3}} {\cal D}^{**}_{0,0} \, , \nonumber \\
X^\prime_0 & : \ & +\frac{1}{2\sqrt{3}} {\cal D}_{0,0} +\frac 1 6 {\cal D}^{**}_{0,0} \, , \nonumber \\
X_1              & : \ & +\frac 1 3 {\cal D}^{*g}_{1,\lambda} \, , \nonumber \\
Z                 & : \ & -\frac{i}{3} {\cal D}^{**}_{1,\lambda} \, , \nonumber \\
Z^\prime    & : \ & +\frac{i}{3} {\cal D}^{*u}_{1,\lambda} \, , \nonumber \\
X_2              & : \ & -\frac{1}{3} {\cal D}^{**}_{2,\lambda} \, .
\label{eq:Overlaps}
\end{eqnarray}
Before exhibiting the explicit mixing matrices $\bm{G}^{\eta, \lambda}$, we note several remarkable features of Eqs.~(\ref{eq:Overlaps}): First, all of the expressions are quite compact: Only one ${\cal D}$ structure overlaps with each $\de\bde$ state, with the exception of the $0^{++}$ states $X_0^{(\prime)}$, and the states that overlap have the same value of $s$.  $X_1$, for example, couples only to $s \! = \! 1$ $D\bar D^*$ states and $Z$ couples only to $s \! = \! 1$ $D^* \! \bar D^*$ states.  If one further combines $X_0^{(\prime)}$ into the following unitarily equivalent basis:
\begin{eqnarray}
+\frac 1 2 X_0 \! - \frac{\sqrt{3}}{2} X^\prime_0 & : \ & -\frac 1 3 {\cal D}_{0,0} \, , \nonumber \\
+\frac{\sqrt{3}}{2} X_0 \! + \frac 1 2 {\tilde X}^\prime_0 & : \ & +\frac 1 3 {\cal D}^{**}_{0,0} \, ,
\end{eqnarray} 
then one sees that every $\de\bde$ state carries the same total weight $\left( \frac 1 3 \right)$ with respect to the basis of ${\cal D}$ combinations.    This specific factor arises through the color Fierz reordering [Eq.~(\ref{eq:ColorFierz})] between $\de\bde$ and di-meson interpolating operators.

Second, the numerical overlap values of Eqs.~(\ref{eq:Overlaps}) hold separately for each allowed value of $\lambda$.  This great simplification will have profound consequences when we compute the actual diabatic potential matrices $\bm{V}(r)$ in the next subsection.

Using Eqs.~(\ref{eq:Overlaps}), we now compile the explicit mixing matrices $\bm{G}^{\eta, \lambda}$.
\begin{equation}
\label{eq:Gg0Matrix}
\bm{G}^{g,0} (r) = \left( \renewcommand{\arraystretch}{1.3} \begin{array}{cc} V_{\de\bde} (r) \, I_4 & \bm{v}^\dagger_{g,0} \, h(r) \\ \bm{v}_{g,0} \, h(r) & {\cal M}_{g,0} \end{array} \right) ,
\end{equation}
where the elements of the rows/columns correspond, in order, to the states:
\begin{equation}
\{ X_{0,0}, \, X^\prime_{0,0}, \, X_{1,0}, \, X_{2,0}, \, {\cal D}_{0,0}, \, {\cal D}^{**}_{0,0}, \, {\cal D}^{*g}_{1,0}, \, {\cal D}^{**}_{2,0} \} ,
\end{equation}
while
\begin{equation}
\label{eq:Mg0Matrix}
I_4 \equiv {\rm diag} \{ 1, 1, 1, 1 \} \, , \ {\cal M}_{g,0} = \frac{\Delta}{2} \, {\rm diag} \{ -3, +1, -1, +1 \} \, ,
\end{equation}
where $\Delta$ is defined in Eq.~(\ref{eq:MassSplit}), and
\begin{equation}
\label{eq:vg0Matrix}
\bm{v}_{g,0} = \left( \renewcommand{\arraystretch}{1.3} \begin{array}{cccc}
 -\frac 1 6 & +\frac{1}{2\sqrt{3}} & 0 & 0 \\ +\frac{1}{2\sqrt{3}} & +\frac 1 6 & 0 & 0 \\ 0 & 0 & +\frac 1 3 & 0 \\ 0 & 0 & 0 & -\frac 1 3 \end{array} \right) .
\end{equation}
Next,
\begin{equation}
\bm{G}^{g,\pm 1} (r) = \left( \renewcommand{\arraystretch}{1.3} \begin{array}{cc} V_{\de\bde} (r) \,  I_2 & \bm{v}^\dagger_{g,\pm 1} \, h(r) \\ \bm{v}_{g,\pm 1} \, h(r) & {\cal M}_{g,\pm 1} \end{array} \right) ,
\end{equation}
where the elements of the rows/columns correspond, in order, to the states:
\begin{equation}
\{ X_{1,\pm 1}, \, X_{2,\pm 1}, \, {\cal D}^{*g}_{1,\pm 1}, \, {\cal D}^{**}_{2,\pm 1} \} ,
\end{equation}
while
\begin{equation}
\label{eq:I2andMg1Matrix}
I_2 \equiv {\rm diag} \{ 1, 1 \} \, , \ {\cal M}_{g,\pm 1} = \frac{\Delta}{2} \, {\rm diag} \{ -1, +1 \} \, ,
\end{equation}
and
\begin{equation}
\label{eq:vg1Matrix}
\bm{v}_{g,\pm 1} = \frac 1 3 \, {\rm diag} \{ +1, -1 \} .
\end{equation}
Next,
\begin{equation}
\bm{G}^{g,\pm 2} (r) = \left( \renewcommand{\arraystretch}{1.3} \begin{array}{cc} V_{\de\bde} (r) & -\frac 1 3 h(r) \\ -\frac 1 3 h(r) & +\frac 1 2 \Delta \end{array} \right) ,
\end{equation}
where the elements of the rows/columns correspond, in order, to the states:
\begin{equation}
\{ X_{2,\pm 2}, \, {\cal D}^{**}_{2,\pm 2} \} .
\end{equation}
Finally, for each value of $\lambda \in \{ +1, 0, -1 \}$,
\begin{equation}
\bm{G}^{u,\lambda} (r) = \left( \renewcommand{\arraystretch}{1.3} \begin{array}{cc} V_{\de\bde} (r) \,  I_2 & \bm{v}^\dagger_{u,\lambda} h(r) \\ \bm{v}_{u,\lambda} h(r) & {\cal M}_{u,\lambda} \end{array} \right) ,
\end{equation}
where the elements of the rows/columns correspond, in order, to the states:
\begin{equation}
\{ Z^\prime_\lambda , \, Z_\lambda , \, {\cal D}^{*u}_{1,\lambda}, \, {\cal D}^{**}_{1,\lambda} \} ,
\end{equation}
while
\begin{equation} \label{eq:MuMatrix}
{\cal M}_{u,\lambda} = \frac{\Delta}{2} \, {\rm diag} \{ -1, +1 \} \, ,
\end{equation}
and
\begin{equation}
\label{eq:vuMatrix}
\bm{v}_{u,\lambda} = \frac i 3 \, {\rm diag} \{ -1, +1 \} .
\end{equation}
The independence of elements $\bm{G}^{\eta , \lambda}(r)$ on the specific value of $\lambda$ is apparent from these explicit forms.

\subsection{The Diabatic Potential Matrix}

The final major step in the derivation of the diabatic potential matrices ${\bm V(r)}$ [Eq.~(\ref{eq:FullV})] is the recognition that the matrices $\bm{G}^{\eta, \lambda}$ are not, in general, the true diabatic potential matrices up through $O(1/m_Q)$.  Since we have explicitly included $O(1/m_Q)$ corrections through the mass splitting $\Delta$ [Eq.~(\ref{eq:MassSplit})], consistency requires that all other $O(1/m_Q)$ corrections must be included.  Since the model assumes that $\de, \bde$ behave as fundamental quasiparticles, spin-dependent corrections to $V_{\de\bde}(r)$ first appear at $O(1/m_Q^2)$, and isospin-breaking corrections are numerically small (and are ignored in this work), leaving the heavy-source kinetic-energy operators as the only other $O(1/m_Q)$ corrections that must be included.\footnote{If both light ($u,d$) and $s$ quarks are included in the same formalism, then $O(m_Q^0)$ SU(3)$_{\rm flavor}$-breaking effects also appear.}

But if the heavy quarks are no longer treated as static, then the strict BO approximation no longer holds.  The $\de\bde$ axis $\hat{\bm{r}}$ can no longer be chosen as a single fixed coordinate direction (say, $\hat{\bm{z}}$), and since the angular-momentum projection $\lambda$ is defined in terms of $\hat{\bm{r}}$ [Eq.~(\ref{eq:lambdadef})], then the individual interaction matrices $\bm{G}^{\eta, \lambda} (r)$ no longer provide a full accounting of the interaction potential for the system.  To incorporate such effects, Ref.~\cite{Bruschini:2023zkb} derives in detail precisely how the diabatic approximation produces the full physical interaction potential $V^{\eta, J}_{i, i^\prime, \ell, \ell^\prime}(r)$ in terms of linear combinations (determined entirely by symmetry) of elements $G^{\eta , \lambda}_{i, i^\prime}(r)$ of $\bm{G}^{\eta, \lambda} (r)$.  With reference to the coupled \schro equations written in terms of the radial wave functions $u(r)$,
\begin{multline}
\hspace{-1.2em} \sum_{i^\prime \ell^\prime} \left[ \delta_{i, i^\prime} \delta_{\ell, \ell^\prime} \frac{\hbar^2}{2m_\de} \! \left( \! -\frac{d^2}{dr^2} + \! \frac{ \ell (\ell + 1)}{r^2} \right) \! + \! V^{\eta , J}_{i, i^\prime \! , \ell, \ell^\prime} (r) \right] \! u^{\eta, J}_{i^\prime \! , \ell^\prime} (r) \\
= E \, u^{\eta, J}_{i , \ell} (r) \, ,
\end{multline}
where $i,i^\prime$ signify specific channels that become coupled to produce the physical state (the fundamental $\de\bde$ states and the di-meson states, in all allowed partial waves labeled by orbital angular momentum $\ell, \ell^\prime$), the explicit relation is:
\begin{multline}
V^{\eta , J}_{i, i^\prime \! , \ell, \ell^\prime} (r) = \\
\sqrt{[\ell][\ell^\prime]} \sum_\lambda 
\left( \! \begin{array}{ccc} s_i                  & \ell               & J \\ \lambda & 0 & -\lambda \end{array} \! \right) \!
\left( \! \begin{array}{ccc} s_{i^\prime} & \ell^\prime & J \\ \lambda & 0 & -\lambda \end{array} \! \right) \!
G^{\eta , \lambda}_{i, i^\prime} (r) ,
\label{eq:GtoV}
\end{multline}
where parentheses indicate Wigner $3j$ symbols, and square brackets denote multiplicities of angular momentum multiplets, {(\it i.e.}, $[\ell] \equiv 2\ell + 1$).

It is important to point out that nothing prevents the simultaneous incorporation of coupling between di-meson (or even di-baryon), $Q\bar Q$, and $\de\bde$ states.  Indeed, such a complex configuration has been considered in scattering solutions of the diabatic dynamical diquark model~\cite{Lebed:2023kbm}, in which $\chi_{c1}(3872)$ was treated as a mixture of $D \bar D^*$, $\de\bde$ in the $X_1$ state, and $c\bar c$ in the conventional (and yet-unobserved) $\chi_{c1}(2P)$ state.  The same generalization can be performed in the current formalism, at the price of using rather more complicated potential mixing matrices.

The case in which only the fundamental states $\de\bde$ appear, however, was noted in the previous section to produce a remarkable simplification compared to the $Q\bar Q$ case: The elements $G^{\eta, \lambda}_{i, i^\prime} (r)$ are all independent of the value of $\lambda$ once $\eta, i, i^\prime$ are chosen, and all of them have $s_i \! = \! s_{i^\prime}$.  The factor $G^{\eta, \lambda}_{i, i^\prime} (r)$ may therefore be taken outside of the sum on $\lambda$ in Eq.~(\ref{eq:GtoV}), and the remaining collection of group-theoretical factors can actually be rearranged into the form of an orthogonality relation of Wigner $3j$ symbols~\cite{Edmonds:1957}:
\begin{equation}
\label{eq:3jOrthog}
\sqrt{[\ell][\ell^\prime]} \sum_\lambda 
\left( \! \begin{array}{ccc} s_i                  & \ell               & J \\ \lambda & 0 & -\lambda \end{array} \! \right) \!
\left( \! \begin{array}{ccc} s_i & \ell^\prime & J \\ \lambda & 0 & -\lambda \end{array} \! \right) = \delta_{\ell , \ell^\prime} \, \Delta \left( s_i , \ell , J \right) ,
\end{equation}
where $\Delta ( j_1 , j_2, j_3)$ is the angular momentum triangle-rule constraint.  It then immediately follows that, for $\de\bde$ states,
\begin{equation}
\label{eq:GtoVDiag}
V^{\eta , J}_{i, i^\prime \! , \ell, \ell^\prime} (r) = \delta_{\ell , \ell^\prime} \, \Delta \left( s_i , \ell , J \right) G^{\eta, \lambda}_{i, i^\prime} (r) ,
\end{equation}
for each allowed value of $s_i \! = \! s_{i^\prime}$ and of $\lambda$.  Notably, no mixed partial waves arise.  This result also applies in the $Q\bar Q$ case~\cite{Bruschini:2023zkb} to the diagonal entries (the fundamental potential $V_{Q\bar Q}(r)$ and the mass-splitting terms), but in the $\de\bde$ case it applies to all channels, and in all allowed partial waves.

The last contribution to be included into the full effective potential is the centrifugal term.  The zero of energy for the system is defined as twice the spin-averaged $D^{(*)}$ mass $m$ [Eq.~(\ref{eq:MassAve})], and any $O(m_Q^0)$ difference between $m$ and the mass of the $\de\bde$ state is incorporated into the constant potential offset $V_0$ in Eq.~(\ref{eq:Cornell}).  The reduced mass for the $\de\bde$ system is $\mu \! = \! m_\de$, and that for the di-meson system (apart from the spin splittings $\propto \! \Delta$ [Eq.~(\ref{eq:MassSplit})]) is the $D^{(*)}$ mass average $m$.  Since the centrifugal potential $\Delta V(r)$ is inversely proportional to mass, both reduced masses can be taken to equal $\frac 1 2 m_\de$, noting that the $O(m_Q^0)$ difference from ${\frac 1 2} m$ in the di-meson case leads to an $O(1/m_Q^2)$ correction.  We then have:
\begin{equation} \label{eq:Centrif}
\Delta V_{i,i^\prime,\ell,\ell^\prime} (r) = +\frac{\ell (\ell + 1)}{m_\de r^2} \, \delta_{i,i^\prime} \delta_{\ell, \ell^\prime} \, .
\end{equation}

We are now ready to consider all partial waves up through those producing total angular momentum $J \! = \! 2$.  We have assumed light d.o.f.\ that create the BO potential in the orbital $\Sigma^+_g(1S)$, which couple to the heavy d.o.f.\ described by the $\de\bde$ states defined in Eqs.~(\ref{eq:Swavediquark}).  While the path leading to this point has been more involved than that for the case of mixing with $Q\bar Q$ states (Ref.~\cite{Bruschini:2023zkb}), the final results in the $\de\bde$ case can be expressed in a much more compact form, owing to the simplifications just derived.

\subsection{Explicit Diabatic Potential Matrices}
\label{sec:ExplicitVs}

In Eqs.~(\ref{eq:Gg0Matrix})-(\ref{eq:vuMatrix}), we note multiple appearances of the same matrix structures [related, of course, to the independence of the mixing matrices $\bm{G}^{\eta, \lambda}(r)$ on $\lambda$].  Let us define:
\begin{equation} \label{eq:v0Matrix}
\bm{v}^{(0)} \equiv \left( \! \renewcommand{\arraystretch}{1.3} \begin{array}{cc} -\frac 1 6 & +\frac{1}{2\sqrt{3}} \\ +\frac{1}{2\sqrt{3}} & +\frac 1 6 \end{array} \! \right) ,
\end{equation}
which equals the upper 2$\times$2 block of $\bm{v}_{g,0}$ from Eq.~(\ref{eq:vg0Matrix}), and
\begin{equation}
{\cal M}^{(0)} \equiv \frac{\Delta}{2} \, {\rm diag} \{ -3, +1 \} \, ,
\end{equation}
which equals the upper 2$\times$2 block of ${\cal M}_{g,0}$ in Eq.~(\ref{eq:Mg0Matrix}).  Once again, $I_4$ and $I_2$ [Eqs.~(\ref{eq:Mg0Matrix}), (\ref{eq:I2andMg1Matrix})] are the 4$\times$4 and 2$\times$2 identity matrices, respectively.  From these, we form
\begin{equation}
\bm{V}^{(g,0)} (r) \equiv \left( \renewcommand{\arraystretch}{1.3} \begin{array}{cc} V_{\de\bde}(r) \, I_2 & \bm{v}^{(0) \dagger} h(r) \\ \bm{v}^{(0)} h(r) & {\cal M}^{(0)} \end{array} \right) .
\end{equation}
The remaining entries of $\bm{G}^{g, {0}} (r)$ and those of $\bm{G}^{g,\pm 1} (r)$ and $\bm{G}^{g,\pm 2} (r)$ naturally generate the matrices:
\begin{equation}
\bm{V}^{(g,1)} (r) \equiv \left( \renewcommand{\arraystretch}{1.3} \begin{array}{cc} V_{\de\bde}(r) & +\frac 1 3 h(r) \\ +\frac 1 3 h(r) & -\frac 1 2 \Delta \end{array} \right) ,
\label{eq:Vg1}
\end{equation}
and
\begin{equation}
\bm{V}^{(g,2)} (r) \equiv \left( \renewcommand{\arraystretch}{1.3} \begin{array}{cc} V_{\de\bde}(r) & -\frac 1 3 h(r) \\ -\frac 1 3 h(r) & +\frac 1 2 \Delta \end{array} \right) ,
\end{equation}
while the entries of $\bm{G}^{u,0} (r)$ and $\bm{G}^{u,\pm 1} (r)$ naturally generate the matrices:
\begin{equation}
\bm{V}^{(u,0)} (r) \equiv \left( \renewcommand{\arraystretch}{1.3} \begin{array}{cc} V_{\de\bde}(r) & -\frac i 3 h(r) \\ +\frac i 3 h(r) & -\frac 1 2 \Delta \end{array} \right) ,
\label{eq:Vu0}
\end{equation}
and
\begin{equation}
\bm{V}^{(u,1)} (r) \equiv \left( \renewcommand{\arraystretch}{1.3} \begin{array}{cc} V_{\de\bde}(r) & +\frac i 3 h(r) \\ -\frac i 3 h(r) & +\frac 1 2 \Delta \end{array} \right) .
\label{eq:Vu1}
\end{equation}
The centrifugal terms can also be abbreviated:
\begin{equation}
\bm{L}_{n} (r) \equiv \frac{\hbar^2}{m_\de r^2} I_n ,
\end{equation}
for $n=2, 4$, and are included in the expressions for $\bm{V}$.

$\boxed{J^{PC} \! = 0^{++}}$ has a 6$\times$6 potential matrix that reduces to one 4$\times$4 $\ell \! = \ell^\prime \! = \! 0$ and one 2$\times$2 $\ell \! = \ell^\prime \! = \! 2$ block:
\begin{equation}
\bm{V}^{0^{++}} \! (r) = \left( \! \renewcommand{\arraystretch}{1.3} \begin{array}{cc} \bm{V}^{(g,0)} (r) & 0 \\ 0 &  \bm{V}^{(g,2)} (r) + 6 \bm{L}_2 (r) \end{array} \! \right) ,
\end{equation}
where the rows/columns of $\bm{V}^{0^{++}} \! (r)$, in order, correspond to the states:
\begin{equation}
\{ X_0, \, X^\prime_0, \, {\cal D}_0, \, {\cal D}^{**}_0 \! , \, X_2, \, {\cal D}^{**}_2 \} .
\end{equation}
Note that the simplicity of these forms relies upon the lack of mixed partial waves ($\ell \! = \ell^\prime$), as well as that the potential couplings are also diagonal in $s_i \! =\! s_{i^\prime}$ eigenvalues (but not necessarily in $i, i^\prime$).

$\boxed{J^{PC} \! = 0^{--}}$ has a 2$\times$2 potential matrix with only $\ell \! = \ell^\prime \! = \! 1$ entries:
\begin{equation}
\bm{V}^{0^{--}} \! (r) = \bm{V}^{(g,1)} {(r)} + 2\bm{L}_2 {(r)} \, ,
\end{equation}
where the rows/columns of $\bm{V}^{0^{--}} \! (r)$, in order, correspond to the states:
\begin{equation}
\{ X_1, \, {\cal D}^{*g}_1 \} .
\end{equation}

$\boxed{J^{PC} \! = 0^{-+}}$ has a 4$\times$4 potential matrix that reduces to two 2$\times$2 blocks, both in the partial wave $\ell \! = \ell^\prime \! = \! 1$:
\begin{equation}
\bm{V}^{0^{-+}} \! (r) = \left( \! \renewcommand{\arraystretch}{1.3} \begin{array}{cc} \bm{V}^{(u,0)}(r) & 0 \\ 0 & \bm{V}^{(u,1)}(r) \end{array} \! \right) + 2\bm{L}_4 (r) \, ,
\end{equation}
where the rows/columns of $\bm{V}^{0^{-+}} \! (r)$, in order, correspond to the states:
\begin{equation}
\{ Z^\prime \! , \, {\cal D}^{*u}_1 \! , \, Z, \, {\cal D}^{**}_1 \} .
\end{equation}

$\boxed{J^{PC} \! = 0^{+-}}$ does not arise in this construction, as it cannot be formed from the ground-state BO potential $\Sigma^+_g$~\cite{Lebed:2017min}.

$\boxed{J^{PC} \! = 1^{++}}$ has a 6$\times$6 potential matrix that reduces to one 2$\times$2 $\ell \! = \ell^\prime \! = \! 0$ and two 2$\times$2 $\ell \! = \ell^\prime \! = \! 2$ blocks:
\begin{eqnarray}
\lefteqn{\bm{V}^{1^{++}} \! (r) =} & & \nonumber \\ & & 
\left( \! \renewcommand{\arraystretch}{1.3} \begin{array}{ccc} \bm{V}^{(g,1)} (r) & 0 & 0 \\ 0 & \bm{V}^{(g,1)} (r) \! + 6\bm{L}_2 (r) & 0 \\ 0 & 0 & \bm{V}^{(g,2)} (r) \! + 6\bm{L}_2 (r) \end{array} \!\right) \! , \nonumber \\
\label{eq:V1++}
\end{eqnarray}
where the rows/columns of $\bm{V}^{1^{++}} \! (r)$, in order, correspond to the states:
\begin{equation}
\{ X_1, \, {\cal D}^{*g}_1 \! , \,  X_1, \, {\cal D}^{*g}_1 \! , \, X_2, \, {\cal D}^{**}_2 \} .
\end{equation}
Note that duplicated states (here, $X_1$ and ${\cal D}^{*g}_1$) refer to distinct partial waves.

$\boxed{J^{PC} \! = 1^{--}}$ has a 10$\times$10 potential matrix, which reduces to one 4$\times$4 and two 2$\times$2 $\ell \! = \ell^\prime \! = \! 1$ blocks, and a 2$\times$2 $\ell \! = \ell^\prime \! = \! 3$ block:
\begin{eqnarray}
\bm{V}^{1^{--}} \! (r) \! & = & \! \left( \renewcommand{\arraystretch}{1.3} \begin{array}{cccc} \bm{V}^{(g,0)} (r) & 0 & 0 & 0 \\ 0 &  \bm{V}^{(g,1)} (r) & 0 & 0 \\ 0 & 0 & \bm{V}^{(g,2)} (r) & 0 \\ 0 & 0 & 0 & \bm{V}^{(g,2)} (r) \end{array} \right)
\nonumber \\
& + & \! \left( \renewcommand{\arraystretch}{1.3} \begin{array}{@{\hspace{0.7em}}c@{\hspace{1.6em}}c@{\hspace{1.6em}}c@{\hspace{1.3em}}c} 2\bm{L}_4 (r) & 0 & 0 & 0 \\ 0 &  2\bm{L}_2 (r) & 0 & 0 \\ 0 & 0 & 2\bm{L}_2 (r) & 0 \\ 0 & 0 & 0 & 12\bm{L}_2 (r) \end{array} \right) \! , \nonumber \\
\end{eqnarray}
where the rows/columns of the matrix $\bm{V}^{1^{--}} \! (r)$, in order, correspond to the states:
\begin{equation}
\{ X_0, \, X^\prime_0, \, {\cal D}_0, \, {\cal D}^{**}_0 \! , \, X_1, \, {\cal D}^{*g}_1 \! , \, X_2, \, {\cal D}^{**}_2 \! , \, X_2, \, {\cal D}^{**}_2 \} .
\end{equation}

$\boxed{J^{PC} \! = 1^{+-}}$ has an 8$\times$8 potential matrix that reduces to two 2$\times$2 $\ell \! = \ell^\prime \! = \! 0$ blocks and two 2$\times$2 $\ell \! = \ell^\prime \! = \! 2$ blocks:
\begin{eqnarray}
\lefteqn{\bm{V}^{1^{+-}} \! (r) =} & & \nonumber \\
& & \left( \! \renewcommand{\arraystretch}{1.3} \begin{array}{cccc} \bm{V}^{(u,0)}(r) & 0 & 0 & 0 \\ 0 &  \bm{V}^{(u,1)}(r) & 0 & 0 \\ 0 & 0 &  \bm{V}^{(u,0)}(r) & 0 \\ 0 & 0 & 0 & \bm{V}^{(u,1)}(r) \end{array} \! \right) \! , \nonumber \\
& + & \left( \! \renewcommand{\arraystretch}{1.3} \begin{array}{@{\hspace{2em}}c@{\hspace{4.1em}}c@{\hspace{2.5em}}c@{\hspace{0.9em}}c} 0 & 0 & 0 & 0 \\ 0 & 0 & 0 & 0 \\ 0 & 0 & + 6\bm{L}_2 (r) & 0 \\ 0 & 0 & 0 & + 6\bm{L}_2 (r) \end{array} \! \right) \! ,
\label{eq:V1+-}
\end{eqnarray}
where the rows/columns of $\bm{V}^{1^{+-}} \! (r)$, in order, correspond to the states:
\begin{equation}
\{ Z^\prime \! , \, {\cal D}^{*u}_1 \! , \, Z, \, {\cal D}^{**}_1 \! , \, Z^\prime \! , \, {\cal D}^{*u}_1 \! , \, Z, \, {\cal D}^{**}_1 \} .
\end{equation}

$\boxed{J^{PC} \! = 1^{-+}}$ has a 4$\times$4 potential matrix that reduces to two 2$\times$2 $\ell \! = \ell^\prime \! = \! 1$ blocks:
\begin{equation}
\bm{V}^{1^{-+}} \! (r) =
\left( \! \renewcommand{\arraystretch}{1.3} \begin{array}{cc} \bm{V}^{(u,0)} (r) & 0 \\ 0 & \bm{V}^{(u,1)} (r) \end{array} \!\right) + 2\bm{L}_4 (r) \, , \end{equation}
where the rows/columns of $\bm{V}^{1^{-+}} \! (r)$, in order, correspond to the states:
\begin{equation}
\{ Z^\prime \! , \, {\cal D}^{*u}_1 \! , \, Z, \, {\cal D}^{**}_1 \} .
\end{equation}

$\boxed{J^{PC} \! = 2^{++}}$ has a 14$\times$14 potential matrix that reduces to two 2$\times$2 $\ell \! = \ell^\prime \! = \! 0$ blocks, one 4$\times$4 and two 2$\times$2 $\ell \! = \ell^\prime \! = \! 2$ blocks, and one 2$\times$2 $\ell \! = \ell^\prime \! = \! 4$ block:
\begin{widetext}
\begin{equation}
\bm{V}^{2^{++}} \! (r) = \left( \! \begin{array}{cccccc} \bm{V}^{(g,1)} (r) & 0 & 0 & 0 & 0 & 0 \\ 0 & \bm{V}^{(g,2)} (r) & 0 & 0 & 0 & 0 \\ 0 & 0 & \bm{V}^{(g,0)} (r) \! + 6\bm{L}_4 (r) & 0 & 0 & 0 \\ 0 & 0 & 0 & \bm{V}^{(g,1)} (r) \! + 6\bm{L}_2 (r) & 0 & 0 \\ 0 & 0 & 0 & 0 & \bm{V}^{(g,2)} (r) \! + 6\bm{L}_2 (r) & 0 \\ 0 & 0 & 0 & 0 & 0 &  \bm{V}^{(g,2)} (r) \! + 20\bm{L}_2 (r)
\end{array} \right) ,
\end{equation}
where the rows/columns of $\bm{V}^{2^{++}} \! (r)$, in order, correspond to the states:
\begin{equation}
\{ X_1, \, {\cal D}^{*g}_1 \! , X_2, \, {\cal D}^{**}_2 \! , \! \, X_0, \, X^\prime_0, \, {\cal D}_0, \, {\cal D}^{**}_0 \! , \, X_1, \, {\cal D}^{*g}_1 \! , \, X_2, \, {\cal D}^{**}_2 \! , \, X_2, \, {\cal D}^{**}_2 \} .
\end{equation}
\end{widetext}

$\boxed{J^{PC} \! = 2^{--}}$ has an 8$\times$8 potential matrix that reduces to two 2$\times$2 $\ell \! = \ell^\prime \! = \! 1$ and two 2$\times$2 $\ell \! = \ell^\prime \! = \! 3$ blocks:
\begin{eqnarray}
\bm{V}^{2^{--}} \! (r) \! & = & \! \left( \renewcommand{\arraystretch}{1.3} \begin{array}{cccc} \bm{V}^{(g,1)} (r) & 0 & 0 & 0 \\ 0 &  \bm{V}^{(g,2)} (r) & 0 & 0 \\ 0 & 0 & \bm{V}^{(g,1)} (r) & 0 \\ 0 & 0 & 0 & \bm{V}^{(g,2)} (r) \end{array} \right)
\nonumber \\
& + & \! \left( \renewcommand{\arraystretch}{1.3} \begin{array}{@{\hspace{0.7em}}c@{\hspace{1.6em}}c@{\hspace{1.3em}}c@{\hspace{1.1em}}c} 2\bm{L}_2 (r) & 0 & 0 & 0 \\ 0 &  2\bm{L}_2 (r) & 0 & 0 \\ 0 & 0 & 12\bm{L}_2 (r) & 0 \\ 0 & 0 & 0 & 12\bm{L}_2 (r) \end{array} \right) \! , \nonumber \\
\end{eqnarray}
where the rows/columns of the matrix $\bm{V}^{2^{--}} \! (r)$, in order, correspond to the states:
\begin{equation}
\{ X_1, \, {\cal D}^{*g}_1 \! , \, X_2, \, {\cal D}^{**}_2 \! , \,  X_1, \, {\cal D}^{*g}_1 \! , \, X_2, \, {\cal D}^{**}_2 \} .
\end{equation}

$\boxed{J^{PC} \! = 2^{-+}}$ has an 8$\times$8 potential matrix that reduces to two 2$\times$2 $\ell \! = \ell^\prime \! = \! 1$ and two 2$\times$2 $\ell \! = \ell^\prime \! = \! 3$ blocks:
\begin{eqnarray}
\bm{V}^{2^{-+}} \! (r) \! & = & \! \left( \renewcommand{\arraystretch}{1.3} \begin{array}{cccc} \bm{V}^{(u,0)} (r) & 0 & 0 & 0 \\ 0 &  \bm{V}^{(u,1)} (r) & 0 & 0 \\ 0 & 0 & \bm{V}^{(u,0)} (r) & 0 \\ 0 & 0 & 0 & \bm{V}^{(u,1)} (r) \end{array} \! \right)
\nonumber \\
& + & \! \left( \renewcommand{\arraystretch}{1.3} \begin{array}{@{\hspace{0.7em}}c@{\hspace{1.7em}}c@{\hspace{1.4em}}c@{\hspace{1.1em}}c} 2\bm{L}_2 (r) & 0 & 0 & 0 \\ 0 &  2\bm{L}_2 (r) & 0 & 0 \\ 0 & 0 & 12\bm{L}_2 (r) & 0 \\ 0 & 0 & 0 & 12\bm{L}_2 (r) \end{array} \right) \!\! ,  \nonumber \\
\end{eqnarray}
where the rows/columns of the matrix $\bm{V}^{2^{-+}} \! (r)$, in order, correspond to the states:
\begin{equation}
\{ Z^\prime \! , \, {\cal D}^{*u}_1 \! , \, Z, \, {\cal D}^{**}_1 \! , \,  Z^\prime \! , \, {\cal D}^{*u}_1 \! , \, Z, \, {\cal D}^{**}_1 \} .
\end{equation}

$\boxed{J^{PC} \! = 2^{+-}}$ has a 4$\times$4 potential matrix that reduces to two 2$\times$2 $\ell \! = \ell^\prime \! = \! 2$ blocks:
\begin{equation}
\bm{V}^{2^{+-}} \! (r) = \left( \! \renewcommand{\arraystretch}{1.3} \begin{array}{cc} \bm{V}^{(u,0)}(r) & 0 \\ 0 & \bm{V}^{(u,1)}(r) \end{array} \! \right) + 6\bm{L}_4 (r) \, ,
\end{equation}
where the rows/columns of $\bm{V}^{2^{+-}} \! (r)$, in order, correspond to the states:
\begin{equation} \label{eq:2+-States}
\{ Z^\prime \! , \, {\cal D}^{*u}_1 \! , \, Z, \, {\cal D}^{**}_1 \} .
\end{equation}

Equations (\ref{eq:v0Matrix})-(\ref{eq:2+-States}) represent the central results of this work.  Despite their cumbersome size in some cases, they exhibit extremely compact and repeatedly occurring substructures, and these simplifications are direct consequences of HQSS\@.  Before concluding this section, let us explore the reasons that the $\de\bde$ case gives much simpler expressions than the $Q\bar Q$ case.

The key simplification is that the $\de\bde$ and $D^{(*)} \bar D^{(*)}$ states in their respective $S$ waves [Eqs.~(\ref{eq:Swavediquark}), (\ref{eq:DiMeson1})-(\ref{eq:DiMeson2})] form a one-to-one correspondence in $J^{PC}$ quantum numbers (with $J \! = \! s_i$ or $s_{i^\prime}$).  Both are $Q\bar Q q\bar q$ states with various symmetry properties differing only in their detailed color structure, and thus both produce the same collection of total $J^{PC}$ values: All have $P \! = \! +$, and the set consists of $0^{++}$ (2 states), $1^{+-}$ (2 states), $1^{++}$, and $2^{++}$.  This correspondence would continue to hold if isospin [or even SU(3)$_{\rm flavor}$] quantum numbers were included.

The specific tabulated $\de\bde$ and $D^{(*)} \bar D^{(*)}$ combinations all have particular symmetry properties, and so it should not be surprising that the overlap results of Eqs.~(\ref{eq:Overlaps}) almost always associate unique $\de\bde$ and $D^{(*)} \bar D^{(*)}$ combinations.  For example, for $J^{PC} \! = \! 1^{+-}$, $Z$ and ${\cal D}^{*u}_1$ both involve combinations of a spin-0 and a spin-1 subcomponent, while $Z^\prime$ and ${\cal D}^{**}_1$ are both spin-1 combinations of two spin-1 subcomponents.  One is thus left with nontrivial mixing only in the $0^{++}$ sector.

Even more generally, however, every case of nontrivial mixing in the $\de\bde$/di-meson system is radically simplified due to $s_i \! = \! s_{i^\prime}$.  We have seen in Sec.~\ref{subsec:MixingMatrices} that the matrix elements $G^{\eta, \lambda}_{i,i^\prime} (r)$ are independent of $\lambda$, which again occurs due to the angular-momentum algebra of both $\de\bde$ and di-meson combinations arising from the same $Q\bar Q q\bar q$ components.  This fact, combined with $s_i \! = \! s_{i^\prime}$, reduces the expression Eq.~(\ref{eq:GtoV}) relating $\bm{G}(r)$ to $\bm{V}(r)$ to a much simpler form [Eq.~(\ref{eq:GtoVDiag})] that not only expresses each element of $\bm{V}(r)$ as a single element of $\bm{G}(r)$ for every allowed partial wave, but also forbids mixed partial waves.

Let us contrast these results with those obtained from $Q\bar Q$ mixing~\cite{Bruschini:2023zkb}.  Here, first of all, the natural ($S$-wave) $Q\bar Q$ parity is $P \! = \! -$, which means that in order to conserve overall parity, the di-meson partial wave $\ell^\prime$ and the $Q\bar Q$ partial wave $\ell$ must differ by an odd integer.  Mixed partial waves are thus necessary and exclusively present in the $Q\bar Q$ case.

The one major simplification of the $Q\bar Q$ case is the conservation of $CP$ quantum number $\eta$, which divides states into separate $\eta \! = \! g$ ($0^{++}$, $1^{++}$, $2^{++}$) and $\eta \! = \! u$ ($1^{+-}$) sectors that do not mix.  In the $Q\bar Q$ case, the $\eta \! = \! g$ states are $s_i \! = \! 1$ and the $\eta \! = \! u$ states are $s_i \! = \! 0$, but the former have nonzero overlaps with di-meson states with each of $s_{i^\prime} \! = \! 0, 1, 2$ (and the latter only with $s_{i^\prime} \! = \! 1$).  The mixing of partial waves and the mixing of spin eigenvalues mean that $Q\bar Q$/di-meson mixing elements of $\bm{V}$ (in contrast to the diagonal $Q\bar Q$ and di-meson elements) are not simplified by the $3j$ symbol orthogonality relation Eq.~(\ref{eq:3jOrthog}).

Lastly, note that all possible $J^{PC}$ quantum numbers (except $0^{{+-}}$) arise in this work (Indeed, simply by having more spin-$\frac 1 2$ constituents, the basic $\de\bde$ states allow for larger values of total spin).  This fact also results from $\de\bde$ and di-meson states having the same $Q\bar Q q\bar q$ components.   In contrast, states formed from $Q\bar Q$ of course forbid the usual exotic $J^{PC}$ values $0^{--}$, $0^{+-}$, $1^{-+}$ , $2^{+-}$, {\it etc.}  

\section{Physical Examples}
\label{sec:Examples}

The linear combinations of the $\de\bde$ states in Eqs.~(\ref{eq:Swavediquark}) carrying definite values of heavy-quark spin $s_{Q\bar Q}$ (as well as light-quark spin $s_{q\bar q}$)~\cite{Lebed:2017min},
\begin{eqnarray}
{\tilde X}_0              & \equiv & \left| 0_{q\bar q} , 0_{Q\bar Q} \right>_0
= +\frac 1 2 X_0 + \frac{\sqrt{3}}{2} X^\prime_0 , \nonumber \\
{\tilde X}^\prime_0 & \equiv & \left| 1_{q\bar q} , 1_{Q\bar Q} \right>_0
= +\frac{\sqrt{3}}{2} X_0 - \frac 1 2 X^\prime_0 , \nonumber \\
{\tilde Z}                  & \equiv & \left| 1_{q\bar q} , 0_{Q\bar Q} \right>_1
= \frac{1}{\sqrt{2}} \left( Z^\prime + Z \right) , \nonumber \\
{\tilde Z}^\prime     & \equiv & \left| 0_{q\bar q} , 1_{Q\bar Q} \right>_1
= \frac{1}{\sqrt{2}} \left( Z^\prime - Z \right) .
\label{eq:X0ZHQspin}
\end{eqnarray}
have overlaps with di-meson states that can be determined immediately from Eq.~(\ref{eq:Overlaps}):
\begin{eqnarray}
{\tilde X}_0              & : & +\frac 1 6 {\cal D}_{0,0} +\frac{1}{2\sqrt{3}} {\cal D}^{**}_{0,0} \, , \nonumber \\
{\tilde X}^\prime_0 & : & -\frac{1}{2\sqrt{3}} {\cal D}_{0,0} +\frac 1 6 {\cal D}^{**}_{0,0} \, , \nonumber \\
{\tilde Z}                  & : & +\frac{i}{3\sqrt{2}} \left( {\cal D}^{*u}_{1,\lambda} - {\cal D}^{**}_{1,\lambda} \right) , \nonumber \\
{\tilde Z}^\prime     & : & +\frac{i}{3\sqrt{2}} \left( {\cal D}^{*u}_{1,\lambda} + {\cal D}^{**}_{1,\lambda} \right) .
\label{eq:HQbasis}
\end{eqnarray}
In addition, the states $X_1$ and $X_2$ are already manifestly eigenstates of heavy-quark spin:
\begin{eqnarray}
X_1 & = & \left| 1_{q\bar q}, 1_{Q\bar Q} \right>_1 \, , \nonumber \\
X_2 & = & \left| 1_{q\bar q}, 1_{Q\bar Q} \right>_2 \, .
\label{eq:X1X2HQspin}
\end{eqnarray}

These combinations are particularly useful if a given exotic state is found to decay overwhelmingly to an $\eta_Q$-like ($0_{Q\bar Q}$) or a $\psi, \Upsilon$-like ($1_{Q\bar Q}$) state.  It is significant to note that should this be the case, then each such state according to Eq.~(\ref{eq:HQbasis}) would couple to two distinct $D^{(*)} \bar D^{(*)}$ structures.  In particular, pure ${\cal D}_{0,0}$ and ${\cal D}^{**}_{0,0}$ states correspond neither to diquark-spin eigenstates nor to eigenstates of heavy-quark spin.

In this section we examine three well-studied~\cite{ParticleDataGroup:2024cfk} hid\-den-charm exotic states: $\chi_{c1}(3872)$ [formerly $X(3872)$], which has $J^{PC} \! = \! 1^{++}$ and $I \! = \! 0$, and $T_{c\bar c 1}(3900)$ and $T_{c\bar c 1}(4020)$, both of which have $J^{PC} \! = \! 1^{+-}$ and $I \! = \! 1$.  In addition, $\chi_{c1}(3872)$ has been observed to decay to $J/\psi$ and $\chi_{c1}$, both of which are $s_{c\bar c} \! = \! 1$ states, but not to $\eta_c$ ($s_{c\bar c} \! = \! 0$); these results agree with the spin content for the state $X_1$ predicted by Eqs.~(\ref{eq:X1X2HQspin}).  Meanwhile, $T_{c\bar c 1}(3900)$ has strong couplings to $J/\psi$ but not $h_c$ (although the decay channel $\eta_c \, \rho$ has been observed), while $T_{c\bar c 1}(4020)$ has strong couplings to $h_c$ but not to $J/\psi$ (nor to $\eta_c \, \rho$).  Meanwhile, $T_{c\bar c 1}(3900)$ decays (dominantly) to $D\bar D^*$ while $D^* \! \bar D^*$ is kinematically forbidden, and $T_{c\bar c 1}(4020)$ decays (dominantly) to $D^* \! \bar D^*$, but the $D \bar D^*$ decay channel has not yet been clearly discerned.  Due to the special significance of $J/\psi$ for $T_{c\bar c 1}(3900)$ and $h_c$ for $T_{c\bar c 1}(4020)$, it is natural to treat the former as a $\tilde Z^\prime$ and the latter as a $\tilde Z$ state according to Eqs.~(\ref{eq:X0ZHQspin}),\footnote{This proposal first appears in Ref.~\cite{Giron:2019cfc}.} and consider the consequences in this formalism.

The $S$-wave block for $J^{PC} \! = \! 1^{++}$ [Eq.~(\ref{eq:V1++}), using Eq.~(\ref{eq:Vg1})] gives
\begin{equation}
\bm{V}^{1^{++}} (r) = \bm{V}^{(g,{1})} (r) = \left( \renewcommand{\arraystretch}{1.3} \begin{array}{cc} V_{\de\bde}(r) & +\frac 1 3 h(r) \\ +\frac 1 3 h(r) & -\frac 1 2 \Delta \end{array} \right) ,
\end{equation}
where the sole di-meson coupling is to the $D \bar D^*$ state ${\cal D}^{*g}_1$, while the $S$-wave blocks for $J^{PC} \! = \! 1^{+-}$ [Eq.~(\ref{eq:V1+-}), using Eqs.~(\ref{eq:Vu0})-(\ref{eq:Vu1})] give
\begin{equation}
\bm{V}^{1^{+-}} (r) = \left( \renewcommand{\arraystretch}{1.3} \begin{array}{cccc} V_{\de\bde} (r) & -\frac i 3 h(r) & 0 & 0 \\ +\frac{i}{3} h(r) & -\frac 1 2 \Delta & 0 & 0 \\ 0 & 0 & V_{\de\bde} (r) & +\frac i 3 h(r) \\ 0 & 0 & -\frac{i}{3} h(r) & +\frac 1 2 \Delta \end{array} \right) \, ,
\end{equation}
where the rows/columns of $\bm{V}^{1^{+-}} \! (r)$, in order, correspond to the states $\{ Z^\prime \! , \, {\cal D}^{*u}_1 \! , \, Z, \, {\cal D}^{**}_1 \}$.  According to Eqs.~(\ref{eq:Overlaps}), the state $Z^\prime$ couples solely to the $D \bar D^*$ combination ${\cal D}^{*u}_1$, and the state $Z$ couples solely to the $D^* \! \bar D^*$ combination ${\cal D}^{**}_1$.  Working instead with the heavy-quark-spin basis states $\tilde Z^\prime, \tilde Z$ [Eqs.~(\ref{eq:HQbasis})] , and defining the equal-admixture $D\bar D^*$-$D^* \! \bar D^*$ states:
\begin{eqnarray}
{\cal D}_1              & \equiv & +\frac{1}{\sqrt{2}} \left( {\cal D}^{*u}_1 \! - {\cal D}^{**}_1 \right) \, , \nonumber \\
{\cal D}^\prime_1 & \equiv & +\frac{1}{\sqrt{2}} \left( {\cal D}^{*u}_1 \! + {\cal D}^{**}_1 \right) \, ,
\end{eqnarray}
then the full potential in the basis $\{ \tilde Z^\prime \! , \, {\cal D}^\prime_1 , \, \tilde Z, \, {\cal D}_1 \}$ reads:
\begin{equation}
\bm{V}^{1^{+-}} (r) = \left( \renewcommand{\arraystretch}{1.3} \begin{array}{cccc} V_{\de\bde} (r) & -\frac i 3 h(r) & 0 & 0 \\ +\frac{i}{3} h(r) & 0 & 0 & -\frac 1 2 \Delta  \\ 0 & 0 & V_{\de\bde} (r) & +\frac i 3 h(r) \\ 0 & -\frac 1 2 \Delta & -\frac{i}{3} h(r) & 0 \end{array} \right) \, ,
\end{equation}
which indicates that both heavy-quark-spin eigenstates $\tilde Z^\prime, \tilde Z$ have equal couplings to $D \bar D^*$ and $D^* \! \bar D^*$ states, and moreover with the same absolute weight for both $\tilde Z^\prime$ and $\tilde Z$ if the $D^*$-$D$ mass splitting $\Delta$ is ignored.  If $T_{c\bar c 1}(3900)$ and $T_{c\bar c 1}(4020)$ are in fact heavy-quark-spin eigenstates, then apart from phase-space constraints, they should have $D\bar D^*$ and $D^* \! \bar D^*$ couplings of equal strength.

Indeed, the magnitude $\frac 1 3 |h(r)|$ of the $D\bar D^*$ (${\cal D}^{*g}_1$) coupling to $X_1$ [identified with $\chi_{c1}(3872)$] equals that for $D\bar D^*$ (${\cal D}^{*u}_1$) to $Z^\prime$, $D^* \! \bar D^*$ (${\cal D}^{**}_1$) to $Z$,  ${\cal D}^\prime_1$ to $\tilde Z^\prime$, and ${\cal D}_1$ to $\tilde Z$.  All of these results are direct consequences of HQSS.

Finally, we note that isospin has so far played no role in this analysis.  While of course fine structure has been incorporated into the model even in its original adiabatic form~\cite{Giron:2019cfc} to compute, for example, mass splittings of $\chi_{c1}(3872)$, $T_{c\bar c 1}(3900)$, and $T_{c\bar c 1}(4020)$, the $I \! = \! 0$ and $I \! = \! 1$ states exhibit a profound difference in the diabatic form: Many of the $I \! = \! 0$ states, $\chi_{c1}(3872)$ included, can mix with $Q\bar Q$ as well as with $\de\bde$ states [$\chi_{c1}(2P)$ in the case of $\chi_{c1}(3872)$].  The full diabatic potential matrix $\bm{V}^{1^{++}}(r)$ in that case must also include direct couplings to $Q\bar Q$ states, including mixed partial waves and to the state ${\cal D}^{**}_2$~\cite{Bruschini:2023zkb}, which is decoupled from the $1^{++}$ state in the pure $\de\bde$ formalism.  Nevertheless, if specific functional forms for both the string-fragmentation amplitude $g(r)$ the quark-rearrangement amplitude $h(r)$ are adopted, then this more complicated analysis remains tractable.  On the other hand, $I \! = \! 1$ states do not mix with $Q\bar Q$ states, and hence the treatment described in this work remains general.

\section{Conclusions} \label{sec:Concl}

In this work we have generalized the diabatic dynamical diquark model, which is designed to incorporate effects that couple di-meson threshold states to fundamen\-tal diquark-antidiquark ($\de\bde$) states.   The extension derived here respects the particular spin couplings and mass-splitting values relevant to each di-meson pair within heavy-quark-spin multiplets (such as $D$ and $D^*$).

This formalism is closely related to that for the diabatic coupling of heavy-quarkonium ($Q\bar Q$) and di-meson threshold states.  However, it presents some key differences, some of which complicate the formalism, and some of which greatly simplify it.  First, one must obtain overlap operators relating the fundamental state ($Q\bar Q$ or $\de\bde$) to particular di-meson operators in specific spin states.  The former case requires one to apply the conventional Fierz reordering to di-meson [$(Q\bar q)(\bar Q q)$] operators with various Dirac structures in order to obtain $(Q\bar Q)(q\bar q)$ structures with the same quantum numbers as $Q\bar Q$ operators, and hence that have nontrivial overlaps.  The latter case requires one to obtain novel Fierz identities relevant to reordering $\de\bde$ operators [$(Qq)(\bar Q \bar q)$] into di-meson [$(Q\bar q)(\bar Q q)$] operators in order to obtain nontrivial overlaps, and is algebraically rather more complicated.  Including $O(1/m_Q)$ heavy-meson mass splittings is straightforward in both cases.

The second key difference requires the fundamental state/di-meson state transition matrices $\bm{G}$ to be converted to the basis in which $O(1/m_Q)$ kinetic-energy operators are diagonalized, thus fully incorporating all corrections to the Born-Oppenheimer heavy-source limit at this order.  In this way, one obtains the full, coupled-channel diabatic potential-energy matrix operator $\bm{V}$ via particular linear combinations of elements of $\bm{G}$.  These linear combinations are not obvious in the $Q\bar Q$ case until they are explicitly derived, and lead to mixing of different partial waves and di-meson states with differing spin eigenvalues.  However, the $\de\bde$ case turns out to be almost trivial because $(Qq)(\bar Q \bar q)$ and $(Q\bar q)(\bar Q q)$ states admit the same set of total $J^{PC}$ values, leading to the individual elements of $\bm{V}$ being directly equal to individual elements of $\bm{G}$, no mixing of partial waves between $\de\bde$ and di-meson states, and no mixing of $\de\bde$ to di-meson states with differing spin eigenvalues.  We have compiled the explicit $\bm{V}$ matrices for all partial waves leading to $J \! = \! 0,1,2$ states.

We also examine a few simple cases directly related to observed states, namely, $\chi_{c1}(3872)$, $T_{c\bar c 1}(3900)$, and $T_{c\bar c 1}(4020)$, and explore how heavy-quark spin symmetry imposes strong constraints that relate these three states.  Future work will numerically explore the detailed relations between these states in the full diabatic formalism, as well as their differences in that the isosinglet $\chi_{c1}(3872)$ can also mix with the $1^{++}$ $c\bar c$ state $\chi_{c1}(2P)$ but the isotriplet $T_{c\bar c 1}$ states cannot mix with $c\bar c$ states.

Other future directions of research include the incorporation of nontrivial interactions between the di-meson pair, a modification that represents a true unification between diquark and molecular di-meson states (and has already accomplished in previous work that uses a universal threshold function for all spin states); the incorporation of fine-structure corrections in both spin and isospin (which has long been studied in the adiabatic dynamical diquark model); and the incorporation of spin-dependent $O(1/m_Q^2)$ effects into each occurrence of the fundamental $\de\bde$ interaction potential appearing in $\bm{V}$.

\begin{acknowledgments}
This work was supported by the National Science Foundation (NSF) under Grant No.\ PHY-2405262.  It contributes to the goals of (but was not funded by) the ExoHad Collaboration, which is supported by the U.S.\ Department of Energy under Grant No.\ DE-SC0023598. 
\end{acknowledgments}

\appendix
\section{Fierz Reorderings} \label{sec:Fierz}

The Fierz identity is simply an expression of the completeness of the Dirac algebra, defined by
\begin{equation}
\left\{ \gamma^\mu , \gamma^\nu \right\} = 2 g^{\mu \nu} I ,
\end{equation}
over the space of $4 \times 4$ matrices, which we express using the standard 16-element basis ($\mu, \nu \in \{ 0, 1, 2, 3 \}$):
\begin{equation} \label{eq:DiracBasis}
\left\{ I, \gamma^\mu, \sigma^{\mu \nu}, \gamma^\mu \gamma_5, i \gamma_5  \right\} ,
\end{equation}
and where we establish the following conventions (consistent with, {\it e.g.}, Ref.~\cite{Peskin:1995ev}):
\begin{eqnarray} \label{eq:DiracDefs}
g^{\mu\nu} & \equiv & {\rm diag} \left\{ +1, -1, -1, -1 \right\} , \nonumber \\
\sigma^{\mu\nu} & \equiv & \frac i 2 \left[ \gamma^\mu , \gamma^\nu \right] , \nonumber \\
\gamma_5 & \equiv & i \gamma^0 \gamma^1 \gamma^2 \gamma^3 , \nonumber \\
\epsilon^{0123} & \equiv & +1 , \nonumber \\
\sigma^{\mu\nu} \gamma_5 & = & +\frac i 2 \epsilon^{\mu\nu\rho\sigma} \sigma_{\rho\sigma} .
\end{eqnarray}
The fermion bilinears containing the basis elements of Eq.~(\ref{eq:DiracBasis}) of course lead to the forms $S,V,T,A,P$, respectively, which are labeled in the usual way according to their properties under Lorentz transformations.

The fundamental Fierz identity of the Dirac algebra, from which all others may be obtained, expresses the outer product of a fermion field $q$ (with Dirac index $\alpha$) and a conjugate fermion field $\bar q^\prime$ (with Dirac index $\beta$):
\begin{widetext}
\begin{equation} \label{eq:ProtoFierz}
q^\alpha \bar q^\prime_\beta = -\frac 1 4 \left\{ \left( \bar q^\prime q \right) \delta^\alpha{}_{\! \beta} + \left( \bar q^\prime \gamma^\mu q \right) \left[ \gamma_\mu \right]^\alpha{}_{\!\! \beta} + \frac 1 2 \left( \bar q^\prime \sigma^{\mu\nu} q \right) \left[ \sigma_{\mu\nu} \right]^\alpha{}_{\!\! \beta} -\left( \bar q^\prime \gamma^\mu \gamma_5 q \right) \left[ \gamma_\mu \gamma_5 \right]^\alpha{}_{\!\! \beta} -\left( \bar q^\prime i \gamma_5 q \right) \left[ i \gamma_5 \right]^\alpha{}_{\!\! \beta} \right\} \, ,
\end{equation}
\end{widetext}
where the factor of $i$ multiplying $\gamma_5$ is included so that the bilinear $\bar q \, i\gamma_5 q$ is Hermitian.  This identity (and each one presented below) holds for fields, and includes the factor of $(-1)$ required by Fermi-Dirac statistics that arises upon the anticommutation of the two fermion fields.  The corresponding identities for spinors are obtained simply by multiplying each of these results by $(-1)$.  The most well-known Fierz reordering, familiar from numerous quantum field theory textbooks, relates 4-fermion operators in which the products of the bilinears transform as Lorentz scalars ($0^+$):
%
\begin{widetext}
\begin{equation} \label{eq:Fierz0}
\renewcommand{\arraystretch}{1.3}
\left(
\begin{array}{c}
\left( \bar q^\prime \vphantom{\bar Q} q \right) \! \left( \bar Q Q \right) \\
\left( \bar q^\prime i \gamma_5 \vphantom{\bar Q} q \right) \! \left( \bar Q i \gamma_5 Q \right) \\
\left( \bar q^\prime \gamma^\mu \vphantom{\bar Q} q \right) \! \left( \bar Q \gamma_\mu Q \right) \\
\left( \bar q^\prime \gamma^\mu \gamma_5 \vphantom{\bar Q} q \right) \! \left( \bar Q \gamma_\mu \gamma_5 Q \right) \\
\left( \bar q^\prime \sigma^{\mu\nu} \vphantom{\bar Q} q \right) \! \left( \bar Q \sigma_{\mu\nu} Q \right) 
\end{array}
\right) = \frac 1 4 \cdot \! \left(
\begin{array}{rrrrr}
 -1 &    +1 &  -1 & +1 &  -\frac 1 2 \\
+1 &     -1 &  -1 & +1 & +\frac 1 2 \\
 -4 &     -4 & +2 & +2 &   0 \\
+4 &   +4 & +2 & +2 &   0 \\
-12 & +12 &   0 &    0 & +2
\end{array}
\right) \left(
\begin{array}{c}
\left( \bar q^\prime \vphantom{\bar Q} Q \right) \! \left( \bar Q q \right) \\
\left( \bar q^\prime i \gamma_5 \vphantom{\bar Q} Q \right) \! \left( \bar Q i \gamma_5 q \right) \\
\left( \bar q^\prime \gamma^\mu \vphantom{\bar Q} Q \right) \! \left( \bar Q \gamma_\mu q \right) \\
\left( \bar q^\prime \gamma^\mu \gamma_5 \vphantom{\bar Q} Q \right) \! \left( \bar Q \gamma_\mu \gamma_5 q \right) \\
\left( \bar q^\prime \sigma^{\mu\nu} \vphantom{\bar Q} Q \right) \! \left( \bar Q \sigma_{\mu\nu} q \right) \end{array}
\right) .
\end{equation}
\end{widetext}
One finds 5 linearly independent such 2-bilinear combinations, in which both bilinears of each term transform under the Lorentz structures conventionally labeled as $S,P,V,A,T$ [in the specific order used in Eq.~(\ref{eq:Fierz0})].  This transformation contains all the Dirac-algebra coefficients required to study components of tetraquark states composed of two $J^P \! = \! 0^+$ diquarks, using the operators defined in Eq.~(\ref{eq:DiquarkOps}).

However, Eq.~(\ref{eq:Fierz0}) is far from being the most general Fierz reordering possible; each of the 4-fermion operators may independently transform as one of $S,V,T,A,P$, leading to a wide variety of Fierz identities.  The most general forms were first tabulated (to our knowledge) in Ref.~\cite{Liao:2012uj}.  The identities presented below may be obtained either by simplifying the rather complex identities of Ref.~\cite{Liao:2012uj}, or by trace methods using the orthogonality of the Dirac basis of Eq.~(\ref{eq:DiracBasis}), or by starting directly with the identity Eq.~(\ref{eq:ProtoFierz}) and simplifying products of Dirac $\gamma$ matrices ({\it e.g.}, using simplifications like the contraction identity $\gamma^\mu \gamma^\nu \gamma_\mu = -2\gamma^\nu$).  We have verified the expressions below through all of these methods.

The numerical matrix $F_0$ in Eq.~(\ref{eq:Fierz0}) (defined to include the coefficient $\frac 1 4$) possesses the property $F_0 \! = \! F_0^{-1}$, which is apparent from the observation that exchanging the fields $Q \leftrightarrow q$ (but leaving $\bar Q$ as is) exchanges the column arrays of 4-fermion operators on the two sides of the equation, thus producing the inverse relation to Eq.~(\ref{eq:Fierz0}).

The corresponding expression for 4-fermion operators whose overall quantum numbers lead to Lorentz transformation as an axial vector ($1^+$) with a single uncontracted index $\mu$ turns out to require an $8 \times 8$ basis:
%
\begin{widetext}
\begin{equation} \label{eq:Fierz1}
\renewcommand{\arraystretch}{1.3}
\left(
\begin{array}{c}
\left( \bar q^\prime \vphantom{\bar Q} q \right) \! \left( \bar Q \gamma_\mu \gamma_5 Q \right) \\
\left( \bar q^\prime \sigma_{\mu\nu} \vphantom{\bar Q} q \right) \! \left( \bar Q \gamma^\nu \gamma_5 Q \right) \\
\left( \bar q^\prime i \gamma_5 \vphantom{\bar Q} q \right) \! \left( \bar Q \gamma_\mu Q \right) \\
\left( \bar q^\prime i \sigma_{\mu\nu} \gamma_5 \vphantom{\bar Q} q \right) \! \left( \bar Q \gamma_\mu \gamma_5 Q \right) \\
\left( \bar q^\prime \gamma_\mu \gamma_5 \vphantom{\bar Q} q \right) \! \left( \bar Q Q \right) \\
\left( \bar q^\prime \gamma^\nu \gamma_5 \vphantom{\bar Q} q \right) \! \left( \bar Q \sigma_{\mu\nu} Q \right) \\
\left( \bar q^\prime \gamma_\mu \vphantom{\bar Q} q \right) \! \left( \bar Q i \gamma_5 Q \right) \\
\left( \bar q^\prime \gamma^\nu \vphantom{\bar Q} q \right) \! \left( \bar Q \, i \sigma_{\mu\nu} \gamma_5 Q \right) 
\end{array}
\right) = \frac 1 4 \cdot \! \left(
\begin{array}{rrrrrrrr}
  -1 &   -i &  +i & -1 &   -1 &  +i &    -i &  -1 \\
+3i &  +i & +3 &  -i &  -3i & +1 &  +3 & +i \\
   -i & +1 &  -1 &  -i &   +i & +1 &   -1 & +i \\
 +3 &   -i & -3i & -1 &  +3 &  +i & +3i & -1 \\
  -1 &  +i &   -i & -1 &   -1 &   -i &   +i & -1 \\
 -3i & +1 & +3 & +i & +3i & +1 &  +3 & -i \\
  +i & +1 &  -1 & +i &    -i & +1 &   -1 & -i \\
 +3 &  +i &+3i & -1 &  +3 &  -i &  -3i  & -1
\end{array}
\right) \left(
\begin{array}{c}
\left( \bar q^\prime \vphantom{\bar Q} Q \right) \! \left( \bar Q \gamma_\mu \gamma_5 q \right) \\
\left( \bar q^\prime \sigma_{\mu\nu} \vphantom{\bar Q} Q \right) \! \left( \bar Q \gamma^\nu \gamma_5 q \right) \\
\left( \bar q^\prime i \gamma_5 \vphantom{\bar Q} Q \right) \! \left( \bar Q \gamma_\mu q \right) \\
\left( \bar q^\prime i \sigma_{\mu\nu} \gamma_5 \vphantom{\bar Q} Q \right) \! \left( \bar Q \gamma^\nu q \right) \\
\left( \bar q^\prime \gamma_\mu \gamma_5 \vphantom{\bar Q} Q \right) \! \left( \bar Q q \right) \\
\left( \bar q^\prime \gamma^\nu \gamma_5 \vphantom{\bar Q} Q \right) \! \left( \bar Q \sigma_{\mu\nu} q \right) \\
\left( \bar q^\prime \gamma_\mu \vphantom{\bar Q} Q \right) \! \left( \bar Q i \gamma_5 q \right) \\
\left( \bar q^\prime \gamma^\nu \vphantom{\bar Q} Q \right) \! \left( \bar Q \, i \sigma_{\mu\nu} \gamma_5 q \right) 
\end{array}
\right) .
\end{equation}
\end{widetext}
The numerical matrix $F_1$ in Eq.~(\ref{eq:Fierz1}) (again defined to include the $\frac 1 4$ coefficient) satisfies $F_1 \! = \! F_1^{-1}$.  This transformation contains all the Dirac-algebra coefficients required to study components of tetraquark states composed of one $J^P \! = \! 0^+$ and one $J^P \! = \! 1^+$ diquark, using the operators defined in Eq.~(\ref{eq:DiquarkOps}).

The Fierz reordering for products of bilinears containing two free Lorentz indices such that their products contain terms transforming as $J^P \! = \! 0^+$, $1^+$ , and $2^+$ [corresponding to components of tetraquark states composed of two $J^P \! = \! 1^+$ diquarks, using the fields defined in Eq.~(\ref{eq:DiquarkOps})] is considerably more complicated.  In this case, one finds 17 distinct 4-fermion structures~\cite{Liao:2012uj}.  However, 5 of these structures carry their free Lorentz indices $\mu,\nu$ entirely through the tensor $g_{\mu\nu}$, and hence the remaining parts of those operators reduce to the forms (with corresponding Fierz reorderings) given in Eq.~(\ref{eq:Fierz0}).  The other 12 operators (generically denoted as $T_{\mu\nu}$), if left in their original forms, mix under Fierz reordering with the initial set of 5 operators.  However, if one subtracts from each of these 12 operators its Lorentz trace $T^\rho{}_{\! \rho}$:
\begin{equation}
T^\prime_{\mu\nu} \equiv T_{\mu\nu} - \frac 1 4 g_{\mu\nu} \, T^\rho{}_{\! \rho} \, , \mbox{ so that } g^{\mu\nu} \, T^\prime_{\mu\nu} = 0 \, ,
\end{equation}
then the remaining set of 12 $T^\prime$ operators decouples from the initial set of 5 operators, and the original 17$\times$17 Fierz reordering matrix is block-diagonalized into a 5$\times$5 and a 12$\times$12 system.  The latter is given by:
%
\begin{widetext}
\begin{equation}
\label{eq:Fierz2}
\renewcommand{\arraystretch}{1.3}
{\bf u} = \frac 1 4 \cdot \! \left(
\begin{array}{rrrrrrrrrrrr}
  -1 &  -1 &   -1 &  -1 &    -i &    -i &   -i &  +i &   +i &   -i & +1 & +1 \\
  -1 &  -1 &   -1 &  -1 &   +i &   +i &  +i &   -i &    -i &  +i & +1 & +1 \\
  -1 &  -1 &   -1 &  -1 &   +i &   +i &   -i &   -i &    -i &   -i &  -1 &  -1 \\
  -1 &  -1 &   -1 &  -1 &    -i &    -i &  +i &  +i &   +i &  +i &  -1 &  -1 \\
  +i &   -i &    -i &  +i &   -1 &  +1 & +1 &  -1 &  +1 &  -1 &  +i &  -i \\
   -i &  +i &   +i &   -i &   -1 &  +1 &  -1 &  -1 &  +1 & +1 &  +i &  -i \\
+2i & -2i & +2i & -2i &  +2 &  +2 &   0 &  +2 & +2 &    0 &   0 &   0 \\
   -i &  +i &   +i &   -i &   -1 &  +1 & +1 &  -1 &  +1 &  -1 &  -i &  +i \\
  +i &   -i &    -i &  +i &   -1 &  +1 &  -1 &  -1 &  +1 & +1 &  -i &  +i \\
+2i & -2i & +2i & -2i &   -2 &   -2 &   0 &   -2 &   -2 &   0 &   0 &   0 \\
 +2 & +2 &   -2 &  -2 &  -2i & +2i &   0 & +2i &  -2i &   0 &   0 &   0 \\
 +2 & +2 &   -2 &  -2 & +2i &  -2i &   0 &  -2i & +2i &   0 &   0 &   0 \\
\end{array}
\right) \left(
\begin{array}{c}
\left( \bar q^\prime \gamma_\mu \vphantom{\bar Q} Q \right) \! \left( \bar Q \gamma_\nu q \right)^\prime \\
\left( \bar q^\prime \gamma_\nu \vphantom{\bar Q} Q \right) \! \left( \bar Q \gamma_\mu q \right)^\prime \\
\left( \bar q^\prime \gamma_\mu \gamma_5 \vphantom{\bar Q} Q \right) \! \left( \bar Q \gamma_\nu \gamma_5 q \right)^\prime \\
\left( \bar q^\prime \gamma_\nu \gamma_5 \vphantom{\bar Q} Q \right) \! \left( \bar Q \gamma_\mu \gamma_5 q \right)^\prime \\
\left( \bar q^\prime \vphantom{\bar Q} Q \right) \! \left( \bar Q \sigma_{\mu\nu} q \right) \\
\left( \bar q^\prime i \gamma_5 \vphantom{\bar Q} Q \right) \! \left( \bar Q \, i \sigma_{\mu\nu} \! \gamma_5 q \right) \\
\!\! \epsilon_{\mu\nu\rho\sigma} \! \left( \bar q^\prime \gamma^\rho \vphantom{\bar Q} Q \right) \!\! \left( \bar Q \gamma^\sigma \! \gamma_5 q \right) \!\! \\
\left( \bar q^\prime \sigma_{\mu\nu} \vphantom{\bar Q} Q \right) \! \left( \bar Q q \right) \\
\left( \bar q^\prime i \sigma_{\mu\nu} \gamma_5 \vphantom{\bar Q} Q \right) \! \left( \bar Q  \, i \gamma_5 q \right) \\
\!\! \epsilon_{\mu\nu\rho\sigma} \! \left( \bar q^\prime \gamma^\rho \gamma_5 \vphantom{\bar Q} Q \right) \!\! \left( \bar Q \gamma^\sigma q \right) \!\! \\
\left( \bar q^\prime \sigma_{\mu\rho} \vphantom{\bar Q} Q \right) \! \left( \bar Q \sigma_\nu{}^\rho q \right)^\prime \\
\left( \bar q^\prime \sigma_{\nu\rho} \vphantom{\bar Q} Q \right) \! \left( \bar Q  \sigma_\mu{}^\rho q \right)^\prime 
\end{array}
\right) ,
\end{equation}
\end{widetext}
where the numerical matrix $F_2$ in Eq.~(\ref{eq:Fierz2}) (again defined to include the coefficient $\frac 1 4$) satisfies $F_2 = F_2^{-1}$, and the elements of the column array {\bf u} are given by the  corresponding operators in the explicit column array in Eq.~(\ref{eq:Fierz2}), upon exchanging the fields $Q \leftrightarrow q$ (but leaving $\bar Q$ as is).

It is important to note that the Fierz reordering expressions of Eqs.~(\ref{eq:Fierz0}), (\ref{eq:Fierz1}), and (\ref{eq:Fierz2}) apply to fermion fields carrying no other indices that require reordering.  In particular, they do not include the effects of the color Fierz reordering necessary for quark-field operators.  Fortunately, for this extension the large matrices obtained above need not be modified, and obtaining the full expressions including color-index reordering requires only minimal modifications.  Here, the crucial expression is the SU(3)$_{\rm color}$ Fierz-reordering identity:
\begin{equation} \label{eq:ColorFierz}
\delta^{\vphantom\dagger}_{k\ell \,} \delta^{\vphantom\dagger}_{jm} = \frac 1 3 \, \delta^{\vphantom\dagger}_{km} \delta^{\vphantom\dagger}_{j\ell} + \frac 1 2 \left( \lambda^a \right)_{km} \left( \lambda^a \right)_{j\ell} ,
\end{equation}
where $\lambda^a$, $a = 1 \ldots 8$, are the usual SU(3)$_{\rm color}$ Gell-Mann matrices.  The full left-hand sides of Eqs.~(\ref{eq:Fierz0}), (\ref{eq:Fierz1}), and (\ref{eq:Fierz2}) are then rewritten as linear combinations of the 4-quark operators given in their original forms, but modified so that their component bilinears now carry color-{\bf 1}/color-{\bf 1} and color-{\bf 8}/color-{\bf 8} indices, with the respective coefficients given in Eq.~(\ref{eq:ColorFierz}).

\section{Charge-Conjugate Fermion Fields} \label{sec:ChargeConj}

We use the conventional definition for the charge-conjugate field $\psi^c$ of an arbitrary fermion field $\psi$:
\begin{equation} \label{eq:psicdef}
\psi^c (x) \equiv C {\bar \psi}^T (x) \, ,
\end{equation}
where $C$ is the charge-conjugation Dirac matrix, which has the (real, antisymmetric) properties:
\begin{equation}
C \gamma^\mu C^{-1} = -\gamma^{\mu \, T} \, , \ \ C^{-1} = C^T = C^\dagger = -C = -C^* \, . 
\end{equation}
In the Weyl basis (used in, {\it e.g.}, Ref.~\cite{Peskin:1995ev}), $C = i\gamma^2 \gamma^0$.  From these relations also follows:
\begin{equation}
{\bar \psi}^c (x) = \psi^T \! (x) \, C \, ,
\end{equation}
as well as the intuitively evident but nontrivial results:
\begin{equation} \label{eq:psicdouble}
\left( \psi^c \right)^c = \psi , \ \ \overline{\psi^c} = {\bar \psi}^c , \ \ \left( {\bar \psi}^c \right)^c = \bar \psi .
\end{equation}
Using Eqs.~(\ref{eq:psicdef})-(\ref{eq:psicdouble}), one also obtains useful bilinear relations for any Dirac fields $\psi (x)$ and $\chi(x^\prime)$:
\begin{eqnarray}
\bar \psi^c (x)  \gamma^\mu \chi(x^\prime) & = &  -\bar \chi^c (x^\prime)   \gamma^\mu \psi (x) , \nonumber \\
\bar \psi^c (x) \, i \gamma_5     \chi(x^\prime) & = & +\bar \chi^c (x^\prime) \, i \gamma_5      \psi (x) ,
\end{eqnarray}
with analogous results for the other Dirac structures.\footnote{The factor $i$ multiplying $\gamma_5$ (and $\sigma_{\mu\nu} \gamma_5$) thus appears in this basis to manifest convenient  properties under Hermitian conjugation.}  We use these expressions in their local ($x^\prime \to x$), and hence gauge-invariant, forms.

\section{Dirac Matrices in the Weyl Spherical-Tensor Basis} \label{subsec:DiracWeyl}

As noted in the previous appendix, the Weyl basis for Dirac matrices~\cite{Peskin:1995ev} is particularly convenient for representing the action of charge conjugation; and as noted in Sec.~\ref{subsec:MixingMatrices}, the spherical-tensor decomposition~\cite{Edmonds:1957} is most convenient for describing particular spin states.  Here we present explicit forms for the relevant Dirac matrices in this combined basis, but also remind the reader that ultimately, our final results can be computed in any basis.

The $\gamma$ matrices in the Weyl basis are given by
\begin{equation} \label{eq:WeylBasis}
\gamma^0 = \left( \! \begin{array}{cc} 0 & 1 \\ 1 & 0 \end{array} \! \right) , \ \ \gamma^i = \left( \!\! \begin{array}{cc} 0 & \sigma^i \\ -\sigma^i & 0 \end{array} \! \right) \, ,
\end{equation}
where $\sigma^i$ are the conventional Pauli matrices.  These definitions may be extended to the spherical basis using $\gamma^\pm$ [Eq.~(\ref{eq:gamma+-})] if one further defines
\begin{equation}
\sigma^\pm \equiv \mp \frac{1}{\sqrt{2}} \left( \sigma^1 \pm i \sigma^2 \right)  = \mp \sqrt{2} \sigma_{\pm, \, {\rm can}} \, ,
\end{equation}
where $\sigma_{\pm, \, {\rm can}}$, the canonical raising/lowering Pauli matrices, satisfy $\sigma_{+, \, {\rm can}} \left| \downarrow \right> \! = \! \left| \uparrow \right>$ and $\sigma_{-, \, {\rm can}} \left| \uparrow \right> \! = \! \left| \downarrow \right>$.  Using this definition and Eqs.~(\ref{eq:DiracDefs}) and (\ref{eq:WeylBasis}), one obtains for $i \! \in \! \left\{ +, -, 3 \right\}$:
\begin{eqnarray}
\gamma_5 = \left( \!\! \begin{array}{cc} -1 & 0 \\ 0 & +1 \end{array} \!\! \right) , & & \ \sigma^{0i} \! = -\sigma^{i0} \! = -i \! \left( \! \begin{array}{cc} \sigma^i & 0 \\ 0 & -\sigma^i \end{array} \!\! \right) , \nonumber \\ \hspace{-2em}
\sigma^{\pm\mp} = \pm i \! \left( \! \begin{array}{cc} \sigma^3 & 0 \\ 0 & \sigma^3 \end{array} \!\! \right) , & & \
\sigma^{\pm 3} \! = -\sigma^{3 \pm} \! = \pm i \! \left( \! \begin{array}{cc} \sigma^{{\pm}} & 0 \\ 0 & \sigma^{{\pm}} \end{array} \! \right) \! .
\end{eqnarray}
The heavy-quark positive/negative-energy projection operators $P_\pm$ [Eq.~(\ref{eq:EProj})] in this basis read:
\begin{equation}
P_\pm = \left( \! \begin{array}{cc} +1 & \pm 1 \\ \mp 1 & +1 \end{array} \! \right) ,
\end{equation}
leading to distinct Dirac-matrix structures having equivalent effects when acting upon heavy-quark fields $Q$ or $\bar Q$.  Explicitly, using 1 for the 4$\times$4 identity matrix and acting upon $Q$ to the right,
\begin{eqnarray}
1 \equiv \gamma^0 , & & \ \gamma_5 \equiv -\gamma^0 \gamma_5 , \nonumber \\
\gamma^\pm \equiv +i\sigma^{0\pm} \equiv \pm i \sigma^{\pm 3} \gamma_5 , & & \ \gamma^3 \equiv +i\sigma^{03} \equiv \pm i \sigma^{\pm\mp} \gamma_5 , \nonumber \\
\gamma^\pm \gamma_5 \equiv -i\sigma^{0\pm} \gamma_5 \equiv \mp i \sigma^{\pm 3} , & & \ \gamma^3 \gamma_5 \equiv -i\sigma^{03} \gamma_5 \equiv \mp i \sigma^{\pm\mp} , \nonumber \\
\label{eq:EquivP+}
\end{eqnarray}
and acting upon $\bar Q$ to the left,
\begin{eqnarray}
1 \equiv -\gamma^0 , & & \ \gamma_5 \equiv -\gamma^0 \gamma_5 , \nonumber \\
\gamma^\pm \equiv +i\sigma^{0\pm} \equiv \pm i \sigma^{\pm 3} \gamma_5 , & & \ \gamma^3 \equiv +i \sigma^{03} \equiv \pm i \sigma^{\pm\mp} \gamma_5 , \nonumber \\
\gamma^\pm \gamma_5 \equiv +i\sigma^{0\pm} \gamma_5 \equiv \pm i \sigma^{\pm 3} , & & \ \gamma^3 \gamma_5 \equiv +i\sigma^{03} \gamma_5 \equiv \pm i \sigma^{\pm\mp} , \nonumber \\
\label{eq:EquivP-}
\end{eqnarray}
but of course, these equivalences hold in any basis.  Note that each corresponding equivalence in Eqs.~(\ref{eq:EquivP+}) and (\ref{eq:EquivP-}) differs at most by a sign. 

\bibliographystyle{apsrev4-2}
\bibliography{diquark}
\end{document}